# Weak Lensing Analysis of A2390 Using Short Exposures

A. Dutta[1], J. R. Peterson[1], T. Rose[2], M. Cianfaglione[3], A. Bonafede[3], G. Li[4], and G. Sembroski[1]
[1] Department of Physics and Astronomy, Purdue University, West Lafayette, IN 47907-2036, USA; dutta26@purdue.edu
[2] Waterloo Center for Astrophysics, University of Waterloo, Waterloo, ON N2L 3G1, Canada
[3] Department of Physics and Astronomy, University of Bologna, Bologna, Italy
[4] Purple Mountain Observatory, West Beijing Road, Nanjing 210008, People's Republic of China

## Abstract

We present a weak lensing analysis of the galaxy cluster A2390 at $z = 0.23$ using second moment shape measurements made in 411 short 60 s exposures. The exposures are obtained in three broadband photometric filters ($g$, $r$, and $i$) using WIYN-ODI. Shape measurement in individual exposures is done using a moment-matching algorithm. Forced measurement is used when the moment-matching algorithm fails to converge at low signal-to-noise ratio. The measurements made in individual images are combined using inverse error weighting to obtain accurate shapes for the sources and hence recover shear. We use PhoSim simulations to validate the shear measurements recovered by our pipeline. We find the mass of A2390 is in agreement with previously published results. We also find the E-mode maps show filamentary structures consistent with baryonic structures and recover most clusters/groups of galaxies found using optical and X-ray data. Thus, we demonstrate the feasibility of using weak lensing to map large-scale structure of the Universe. We also find the central portion of the cluster has a bimodal mass distribution and the relative orientation of the peaks is similar to X-ray. We discuss earlier research on this galaxy cluster, and show that a late-stage merger accounts for all the observed data.

*Unified Astronomy Thesaurus concepts:* Weak gravitational lensing (1797); Galaxy clusters (584)

## 1. Introduction

Galaxy clusters are the largest bound structures in the Universe containing a few hundred to a few thousand individual galaxies. They play an important role in our understanding of the Universe and hierarchical structure formation (W. H. Press & P. Schechter 1974; J. R. Bond et al. 1991; C. Lacey & S. Cole 1993; A. V. Kravtsov & S. Borgani 2012). Mass profiles of galaxy clusters are fundamental to understanding structure formation and constraining cosmological models (G. Holder et al. 2001; C. Cunha et al. 2009). However, the measurement of this is made challenging by the fact that only about 5% of the total mass of galaxy clusters emits light in the visible spectrum; about 15% of the mass emits light in the form of ionized X-rays, and the remaining 80% is expected to be in the form of dark matter (DM; M. Fukugita et al. 1998).

Different methods of measuring the masses of galaxy clusters exist. X-ray observations of the hot ionized intracluster medium (ICM) are expected to trace the general mass profile of a cluster. This method has been widely used along with assumptions of hydrostatic equilibrium and/or a mass profile such as Navarro–Frenk–White (NFW; J. F. Navarro et al. 1995; A. E. Evrard et al. 1996; T. H. Reiprich & H. Böhringer 2002; S. Ettori et al. 2013) to measure the mass of galaxy clusters. These assumptions limit our ability to make accurate measurements since clusters are always in a dynamical state (R. Piffaretti & R. Valdarnini 2008). These include cooling flows and active galactic nuclei (AGN) feedback (A. C. Fabian 1994; M. Gitti et al. 2012), mergers with other clusters, and smaller infalling groups. This is expected since clusters lie at the major intersections of the filamentary structure of the Universe (M. van Haarlem & R. van de Weygaert 1993). Another method of measuring cluster masses is with the velocity dispersion of the cluster members along with the assumption of virial equilibrium (F. Zwicky 1933, 1937; S. Smith 1936; L. Danese et al. 1980). However, a significant majority of the clusters are in a dynamical state where the assumption of virial equilibrium might not hold. The Sunyaev–Zeldovich (SZ) effect (R. A. Sunyaev & Y. B. Zeldovich 1970, 1972, 1980; J. E. Carlstrom et al. 2002) has also been used to measure the mass of galaxy clusters. However, this approach is limited by the high quality of data required, mass-observable scaling relation, and dynamical state of the cluster (E. Krause et al. 2011; J. P. Dietrich et al. 2018).

Weak lensing provides another independent method to measure mass. In weak lensing, coherent distortions in background galaxies (J. A. Tyson et al. 1990; N. Kaiser et al. 1995; G. Squires et al. 1996) are used to infer mass. Unlike other methods, weak lensing makes no assumption about the state of the cluster and directly probes mass. It is equally sensitive to both baryonic matter and DM, making it an ideal tool to directly probe DM distribution in the Universe. Several excellent weak lensing studies on galaxy clusters (M. Postman et al. 2012; A. von der Linden et al. 2014; S. Fu et al. 2022) including A2390 (G. Squires et al. 1996; K. Umetsu et al. 2009; N. Okabe et al. 2010) have been conducted. These have shown that the masses inferred from weak lensing are consistent with X-ray mass estimates, especially for cool core clusters. For merging or disturbed clusters, the mass distributions inferred from X-ray and weak lensing are significantly different (D. Clowe et al. 2006). In such cases, weak lensing measurements of mass profiles are expected to be more accurate. These differences between baryonic and DM interactions during mergers can eventually be used to understand and put constraints on the DM cross section.

In weak lensing, measuring the shapes of galaxies accurately is of great importance. Traditionally, weak lensing studies have used shape measurements of galaxies made in the coadded image. However, images are obtained under a wide range of seeing conditions, cloud cover, and background brightness. Depending on the science requirements, different types of weighted coadd (B. Zackay & E. O. Ofek 2017) are necessary for optimal measurements. It has been suggested that measuring sources in individual images would lead to better measurements since individual images can be weighted optimally to extract better shape information (J. A. Tyson et al. 2008; M. J. Jee & J. A. Tyson 2011; R. Mandelbaum 2018). Using measurements made in individual exposures avoids information lost in the process of coaddition and allows for small-spatial-scale photometry and point-spread function (PSF) variation to be taken into account. This allows for better PSF correction, which in turn leads to better shear recovery. This idea has been used in more recent weak lensing studies such as L. Miller et al. (2013) and J. Zuntz et al. (2018). However, measuring sources in individual images is challenging since a vast majority of the sources are extremely faint and undetectable. We use forced measurement, a generalization of forced photometry proposed by A. Dutta et al. (2024), to measure such low signal-to-noise ratio (SNR) sources. Forced measurement uses reasonable initial guess values of the photon counts, shape, and size of the source being measured to produce values of photon counts, shape, and size that are statistically closer to the true values. The advantage of forced measurement is that it allows for the measurement of sources with SNRs as low as 0.1. This allows us to measure all sources in individual images.

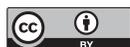

In this paper, we perform a weak lensing analysis of the galaxy cluster A2390, a rich and massive galaxy cluster at $z = 0.23$ (R. G. Abraham et al. 1996). It is a bright X-ray source and shows several arcs due to strong lensing. A2390 has been extensively studied in X-ray (M. Pierre et al. 1996; S. Allen et al. 2001; R. Martino et al. 2014; S. S. Sonkamble et al. 2015), optical (R. G. Abraham et al. 1996; J. B. Hutchings et al. 2002), radio (M. Bacchi et al. 2003; P. Augusto et al. 2006; M. W. Sommer et al. 2016; F. Savini et al. 2019), weak lensing (G. Squires et al. 1996; K. Umetsu et al. 2009; N. Okabe et al. 2010; A. von der Linden et al. 2014), and strong lensing studies (D. Narasimha & S. M. Chitre 1993). In X-ray studies (S. Allen et al. 2001), A2390 shows hints that it has not relaxed since its last merger. Radio studies such as T. Rose et al. (2024a) and M. W. Sommer et al. (2016) have found evidence of gravitational disturbance and turbulence in the ICM, respectively. Weak lensing studies have not reported any signs of a merger.

The paper is organized as follows. In Section 2, we describe our data and the processing pipeline. We describe strategies for coadding our data to maximize source detection. We then measure the detected sources using a moment-matching algorithm and present a novel Monte Carlo PSF correction scheme. We describe in detail how the forced measurement algorithm is used to measure sources in individual exposures. In Section 3, we use PhoSim (J. R. Peterson et al. 2015, 2019, 2020, 2024; C. J. Burke et al. 2019) to simulate 60 short 1 minute exposures to show our pipeline recovers shear accurately. Two different cases with different values of input shear are tested. In Section 4, we present our results and compare them with previous X-ray and radio data. We show the mass structures we recover from weak lensing correspond to an excess of light density in most regions. We also present a scheme to perform 3D mass reconstruction. In Section 5, we discuss in detail the previous studies and their results conducted on A2390 using X-ray, radio, and optical data. Finally, in Section 6, we show that a late-stage merger scenario provides a reasonable explanation for all observed data.

## 2. Observational Data and Processing Pipeline

### 2.1. Data Acquisition

The data for this analysis were obtained using the WIYN 3.5 m telescope at Kitt Peak in Arizona. The diameter of the primary mirror is 3.5 m. For this study, we use the ODI instrument (D. R. Harbeck et al. 2014, 2018), which has a wide field of view of approximately $40' \times 48'$. The focal plane of the instrument is populated with 30 orthogonal-transfer array (OTA) CCDs in a $5 \times 6$ configuration and has a pixel scale of $0\rlap{.}''11$ pixel$^{-1}$. OTAs are a kind of CCD where charges can be moved orthogonally during exposure to increase image sharpness (D. R. Harbeck et al. 2018). However, this feature was not available during our imaging sessions. Each OTA is made of $8 \times 8$ pixel cells. Each pixel cell is composed of $480 \times 496$ pixels. In this document, we use chips to refer to pixel cells. We record images in five separate wide-band photometric filters, $u$, $g$, $r$, $i$, and $z$, over six observing runs from 2017 to 2023. The wide-band photometric filters of ODI are very similar to those used by other survey telescopes such as the Sloan Digital Sky Survey (SDSS; J. E. Gunn & D. H. Weinberg 1994), Pan-STARRS (N. Kaiser et al. 2002), and the Dark Energy Survey (DES; The Dark Energy Survey Collaboration 2005). The data obtained in the 2022 observing run were unfortunately lost due to corrupt header and metadata values. The issue is likely linked to the damage caused by the forest fires on Kitt Peak in the fall of 2022. In total, we lost 90 images, each of 60 s exposure, due to this issue. We also note one OTA, specifically OTA12, was unavailable/nonfunctional in 2023. Over the course of 2017–2023 a few CCD segments became nonfunctional as well. In Table 1, we show the total number of exposures, the PSF size of the coadd, and the depth of the coadd in each filter. The PSF size is defined as the median size of stars obtained using the moment-matching algorithm described in A. Dutta et al. (2024). The depth is defined as the AB magnitude brightness of the finest source detected by SExtractor after the cuts mentioned in Section 2. We also show the number of exposures obtained in each filter in each year.

Rejecting the bad quality/unusable data, seeing conditions ranged from excellent ($0\rlap{.}''6$) to poor ($3''$). The poorest seeing data was obtained in the 2023 observing run, which had a median seeing of $2''$. The normalized histogram of seeing in all images for the $i$, $r$, $g$, and $z$ bands is shown in Figure 1. The median seeing in the $g$, $r$, and $i$ bands is around $1\rlap{.}''0$. We used the default nine-point dithering pattern to fill the CCD gaps. In this process, the telescope pointing is shifted slightly by a predetermined amount nine times to fill gaps between CCD and to ensure uniform coverage of the field. The exposure time for each image is 60 s. This exposure time is reached as a compromise between the short exposure time required for weak lensing (a few seconds ideally) and the 40–50 s readout time for ODI detectors (C. Chang et al. 2012; J. R. Peterson et al. 2015). Short exposures are important for weak lensing since it allows one to reject or de-weight periods of bad seeing. Overall, we have 720 minutes of data divided roughly equally in each

**Table 1**
Observing Runs

| Photometric Band | Exposure Time (s) | PSF Size (pixels) | Depth (mag) | 2017 | 2019 | 2020 | 2021 | 2023 |
|---|---|---|---|---|---|---|---|---|
| u | 120 × 60 | 3.6 | 25 | 22 | 45 | 18 | 17 | 18 |
| g | 161 × 60 | 3.0 | 26 | 36 | 45 | 35 | 27 | 18 |
| r | 151 × 60 | 2.9 | 25.5 | 34 | 45 | 27 | 27 | 18 |
| i | 146 × 60 | 3.0 | 25.5 | 33 | 43 | 27 | 25 | 18 |
| z | 141 × 60 | 2.9 | 26 | 35 | 44 | 26 | 18 | 18 |

**Note.** Column (1): photometric filter. Column (2): corresponding exposure time. Column (3): PSF size of coadd; this is defined as the median size of stars when they are measured using the moment-matching algorithm described in A. Dutta et al. (2024). Column (4): approximate depth of the coadd; the depth is defined as the AB magnitude brightness of the faintest source detected by SExtractor after the cuts mentioned in Section 2. The five right columns show the number of exposures in each filter obtained in the corresponding year. The missing years, i.e., 2018 and 2022, correspond to time lost due to bad weather and technical issues. All exposures are 60 s long.

photometric band. The exposures obtained in 2017, 2020, and 2021 were on gray nights. For 2019, the exposures were obtained on a bright night and hence show significantly high background. In 2023, exposures were obtained on a dark night. We visually examined all images and rejected 36 images where satellite trail or wind shake affected the image significantly.

### 2.2. Correction of Systematics and Defects

All raw data are processed using QuickReduce (R. Kotulla 2013). QuickReduce, the default WIYN data-processing pipeline, corrects for common systematics such as saturation and nonlinearity, persistency effects, dark subtraction, flat-fielding, fringing, removal of cosmic rays, and photometric calibration. Photometric calibration is done using the Pan-STARRS (N. Kaiser et al. 2002; K. C. Chambers et al. 2016) catalog for the g, r, i, and z bands, while the u band is calibrated using the SDSS (J. E. Gunn & D. H. Weinberg 1994) catalog. A World Coordinate System (WCS) match with the Gaia (A. G. A. Brown et al. 2016) catalog is performed to obtain the final calibrated images. However, QuickReduce is not able to perfectly correct for systematics due to small but noticeable degradation of the CCDs. As mentioned previously, a few CCD segments and one OTA gradually degraded and eventually became nonfunctional over the course of 2017–2023. This changing nature of the detector properties is hard to correct for. This was found to be true for the QuickReduce pipeline as well. This caused noticeable amplifier glow, non-Gaussian background, unusually high or low pixel values, and crosstalk in the calibrated images. Below, we discuss each of these issues in detail and how we correct for them.

During visual examination of the calibrated images, it was found some chips had correlated noise. An example of this is shown in Figure 2. Such correlated noise leads to several spurious sources in the resulting coadd when employing SExtractor (E. Bertin & S. Arnouts 1996) for source detection. This limits our ability to detect faint sources. We decided to de-weight these chips during the coadd process instead of completely rejecting them. We use a weighting scheme similar to J. Annis et al. (2014):

$$W = 100 \frac{10^{Z-25}}{(S\sigma_b)^2}, \quad (1)$$

where $Z$ is the zero-point (ZP) of the image, $W$ denotes weight, the average seeing in an image is denoted by $S$, and $\sigma_b$ is the background standard deviation calculated using the $k=3\sigma$ clipping algorithm in astropy (Astropy Collaboration et al. 2013). This weighting scheme de-weights the chips with correlated noise since those chips have higher background variance. We note that we do not completely understand the origin of such noise, but this may simply be a form of pink noise from the electronics (C. M. Hirata et al. 2024).

It was noticed that some images show significant crosstalk patterns. Due to this, pixel values are significantly higher or lower than the background in certain chips adjacent to bright sources. An example of this is shown in Figure 2(b). To correct for the lower pixel values, we find the lower of six standard deviations smaller than the background median and zero, and reject any pixels below this limit. The higher pixel values are harder to correct for since they appear intermittently and mimic real sources. Coadding, described below, was able to get rid of most effects due to crosstalk. Any remaining effects are visually identified in the coadd and those regions are masked during source detection. It was also noticed some chips have background variation significantly higher than expected from Poisson statistics. Most of these chips have imperfections and limit our ability to detect extremely faint sources in the coadd. The problem is severe enough that the weighting scheme and cuts presented above were not able to effectively de-weight these chips. To exclude these chips, we perform cuts in the background median versus variance space. We use the median instead of the mean because it was found to be much more stable. A graph showing the distribution of median versus variance of the background for all chips in the g, r, and i bands is shown in Figure 3. Both background median and variance are obtained after $k=3\sigma$ clipping. We consider the condition

$$0.9 \times \text{Median} \leqslant \text{Variance} \leqslant \text{Median} + 300, \quad (2)$$

where "Variance" is the background variance and "Median" is the background median obtained after $k=3\sigma$ clipping. If a chip violates this condition in more than half the images, the chip is flagged as problematic and its weight is set to zero for all images. This condition effectively rejects the problematic chips.

### 2.3. Coadding and Source Detection

The coadds were performed for each of the five photometric bands separately using the software SWarp (E. Bertin et al. 2002). The SWarp settings used are listed in Table 2. The i- and r-band images were coadded further. A color image of the entire field is shown in Figure 4. The i + r coadd image was inspected visually for any additional imperfections. A few regions of defects were visually identified. Most of these defects come from clumps of unusually high pixel values. We do not understand the origins of these pixels, but they tend to appear at the chip edges or in chips with pre-existing defects. There is also a small amount of crosstalk noise from a few extremely bright stars in the field. We mask out these regions to limit spurious detection. The total percentage of area masked out was less than 2%. This is the final image used for source detection. We employed SExtractor (E. Bertin & S. Arnouts 1996) to detect sources. The SExtractor settings used are shown in Table 2. These settings were found to give us the maximum number of sources without contaminating the sample with too many spurious detections. The details of the number of images coadded and the depth and PSF size of the final image are shown in Table 1. The final number of sources detected in the $i + r$ coadded image is 54,059 and the density is $\sim$31.8 arcmin$^{-2}$.

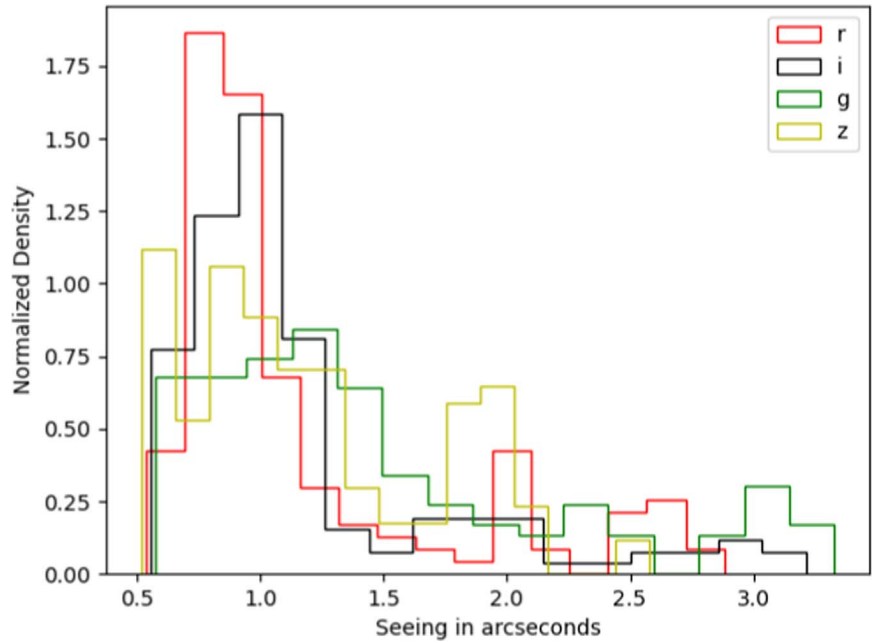

**Figure 1.** Normalized histogram of seeing (in arcseconds) for 151 *r*-band, 146 *i*-band, 161 *g*-band, and 141 *z*-band images. We excluded the *u* band since the image quality is extremely poor and only a few stars are detectable in the entire image. The seeing values are the FWHM size of the stars as calculated by QuickReduce.

### 2.4. Coadd Measurement and Source Classification

The sources detected in the previous step are measured using the moment-matching algorithm developed originally by N. Kaiser et al. (1995) and later modified by G. M. Bernstein & M. Jarvis (2002). We use an adaptive moment-matching algorithm with elliptical Gaussian weights. The shape and size of the weight is the best-fit elliptical Gaussian for the source being measured. The weighted second moments are calculated as

$$Q_{ij} = \int \theta_i \theta_j f(\theta_i, \theta_j) W(\theta_i, \theta_j) d\theta_i d\theta_j, \quad (3)$$

where $\theta$ represents generalized coordinates, $W(\theta_i, \theta_j)$ is the weight function, and $f(\theta_i, \theta_j)$ is the cutout of the source. The polarization or ellipticity parameters are then defined as

$$e_1 = \frac{Q_{11} - Q_{22}}{Q_{11} + Q_{22}} \qquad e_2 = \frac{2Q_{12}}{Q_{11} + Q_{22}}, \quad (4)$$

where $e_1$ represents elongation along the *x*- (positive values) and *y*-axes (negative values), and $e_2$ represents elongation along the $y = x$ or $y = -x$ lines. Traditionally, the quantities $Q_{11}$, $Q_{22}$, and $Q_{12}$ are referred to as $\sigma_{xx}^2$, $\sigma_{yy}^2$, and $\sigma_{xy}^2$, respectively. This algorithm was modified to produce the photon counts, centroid, and size of the source along with ellipticity. We chose the best-fit elliptical Gaussian as our weight function. For a detailed description of this algorithm, see A. Dutta et al. (2024).

The measurements are done in the $i + r$ coadd image. In the first step, the optimal cutout size of each source was determined. To estimate the optimal cutout size, a rough guess of size from SExtractor, $S_{\text{sextractor}}$, is made:

$$S_{\text{sextractor}} = \sqrt{X2\_\text{IMAGE} + Y2\_\text{IMAGE}}, \quad (5)$$

where $X2\_\text{IMAGE}$ is the variance along *x* and similarly $Y2\_\text{IMAGE}$ is the variance along *y*. The second-order moments are obtained from the SExtractor output file. This estimated size is set at 25% larger than the SExtractor size. The lower limit of the estimated size was capped at 4 pixels. Next, a square cutout with side length 8 times the estimated size is made to measure sources. It was found that in rare cases the moment-matching algorithm fails when it is run on such a cutout. However, the algorithm successfully converges when a smaller cutout is made. This happens typically due to light contamination from other nearby sources, leading to gross misestimation of background. For cases where the algorithm initially fails, we reduce the size of the cutout by 2 pixels along the *x*-axis and 2 pixels in the *y*-axis symmetrically from each side. This is repeated until the algorithm successfully converges or the cutout is less than or equal to 24 × 24 pixels. This is the minimum cutout size needed to measure a source. Using this method we obtain a good estimate of the optimal cutout size.

Next, we attempt to reduce the effect of light contamination from nearby sources, thus increasing the accuracy of our measurements. First, we calculate the bound flag (`bndflag`) using the previous measurements. Flags are calculated using the method outlined in Section 2.6. If the bound flag is raised, it indicates that there is a significant chance that the background estimation might be affected by nearby sources, which in turn affects the flux estimation. To correct for this, we use two strategies:

1. The cutout is moved by 3 pixels in a direction 180° away from the nearest source.
2. During the iterative process of moment matching, the background is fixed. To determine the value of background we select three $4 \times 4$ pixel regions. These regions are $s/\sqrt{2}$ away from the central pixel of the cutout, where $s$ is the side length of the cutout. The three regions selected are 90°, 180°, and 270° away from the line joining the source to its nearest source. We also create a $4 \times 4$ array of zeros. The array of zeros was to ensure the background value does not wander too far from the truth, which is very close to zero in the coadds. The median value of these 64 pixels in four $4 \times 4$ cutouts is the fixed background value.

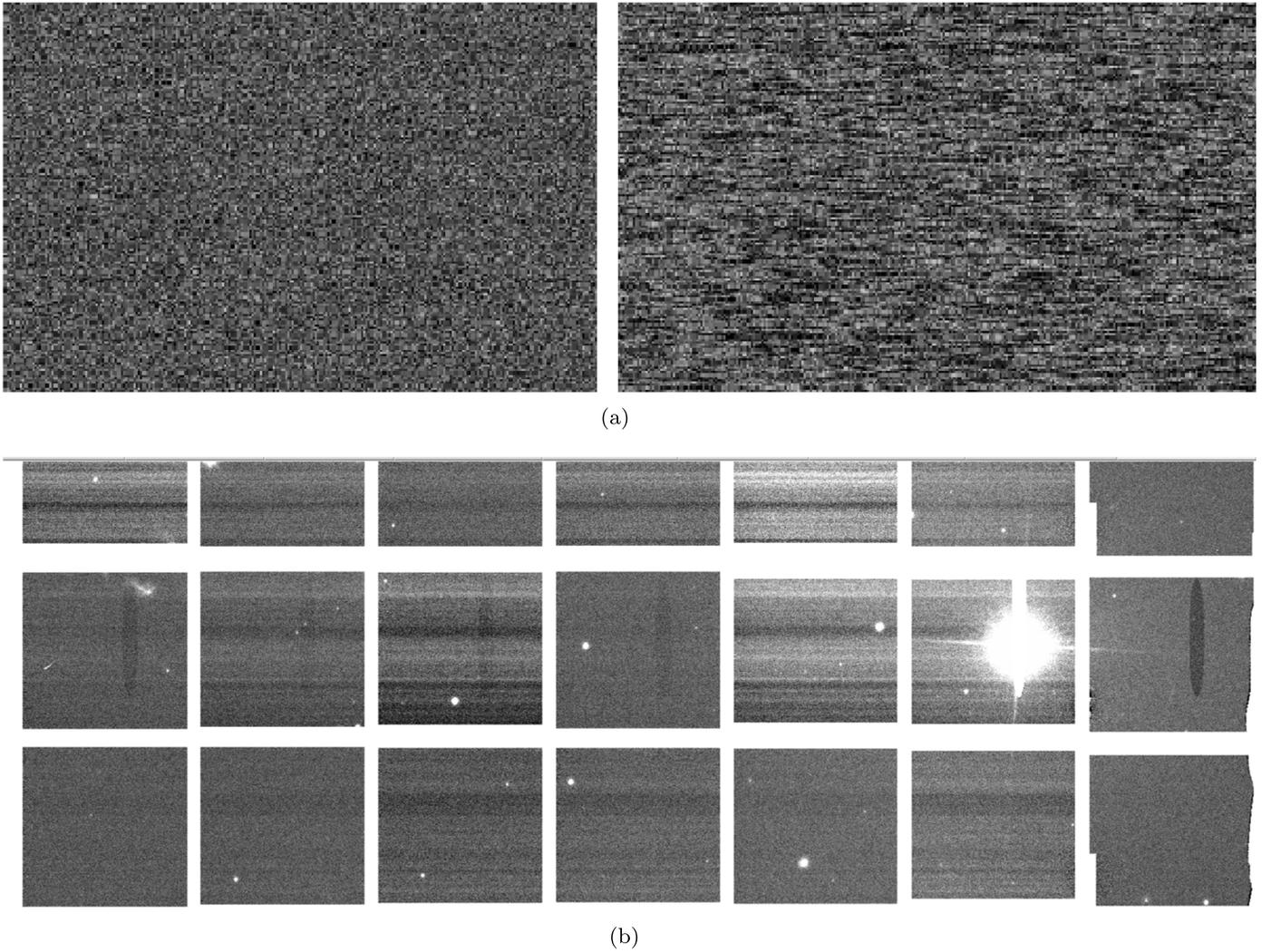

**Figure 2.** (a) On the left an example of background from a typical chip is shown. On the right, the background from chips with correlated noise is shown. Correlated noise gives rise to spurious sources, especially at the faint end. (b) An example of crosstalk caused by very bright sources. This image shows a portion of the OTA. QuickReduce is unable to get rid of all cases of crosstalk arising from extremely bright sources. Coadding reduces them to a nondetectable level. However, some of the severe cases are apparent in the coadd. The vertical line of NaNs through the bright source is evident.

If the bound flag is not raised, we shift the cutout 3 pixels away from the nearest source and remeasure it using the moment-matching algorithm. It was found that the amount of shift is not important as long as the shift is small compared to the size of the cutout. We repeat the same measurement process for the coadd in each of the five photometric bands. The flux versus size graph from the $i + r$ coadd band measurements for all sources is shown in Figure 5. The vertical column feature arises from stars which have approximately fixed size (size of the PSF) but vary in magnitude. Having a clean sample of stars from which to measure the PSF is the first step in accurate PSF estimation. We cross-referenced all sources in our field with the star catalog from Gaia Early Data Release 3 (EDR3; A. G. A. Brown et al. 2021). The criterion used in matching was for the difference in the WCS coordinates of the source in our catalog and Gaia catalog be less than 0″.72. The sources that were matched are shown as red points in Figure 5(a). This gives us an extremely clean sample of stars with which to determine the PSF.

While stars are useful for determining the PSF, they are not useful for weak lensing purposes and only serve to dilute the shear signal. Hence, an important aspect of any weak lensing pipeline is accurately rejecting stars. In our case, rejecting only the stars cross-matched with Gaia EDR3 is not enough. Gaia EDR3 goes to a depth of magnitude 21 in the $g$ band, while our $i + r$ coadded image has a depth of magnitude 26. The mismatch in the depth of the two catalogs means that several fainter stars in our field cannot be cross-matched with the Gaia catalogs. Hence, we visually identified the vertical star strip shown as a black box in Figure 5(a). Any sources inside the black box were rejected from weak lensing analysis. We note that this method is not perfect and some fainter stars outside the black bounding box enter our galaxy sample. To reject extremely bright sources that show the brighter-fatter effect, any sources brighter than magnitude 15, shown by the horizontal yellow line, were rejected. We also reject any sources with size $\geqslant 12$ pixels. These most likely arise from severe blending. This value was selected somewhat arbitrarily and changing this limit does not significantly affect our analysis. We also wish to reject sources that have size $3\epsilon_{\rm PSF}$ below the median PSF value, where $\epsilon_{\rm PSF}$ is the variation of PSF in the image. This is shown by the yellow line on the left in Figure 5(a). All sources left of this yellow line are rejected. A method to determine $\epsilon_{\rm PSF}$ is described in the next section. These are likely spurious sources arising from cosmic rays or defective pixels. After these cuts, we have 45,022 sources remaining, corresponding to a source density of 26.5 arcmin$^{-2}$.

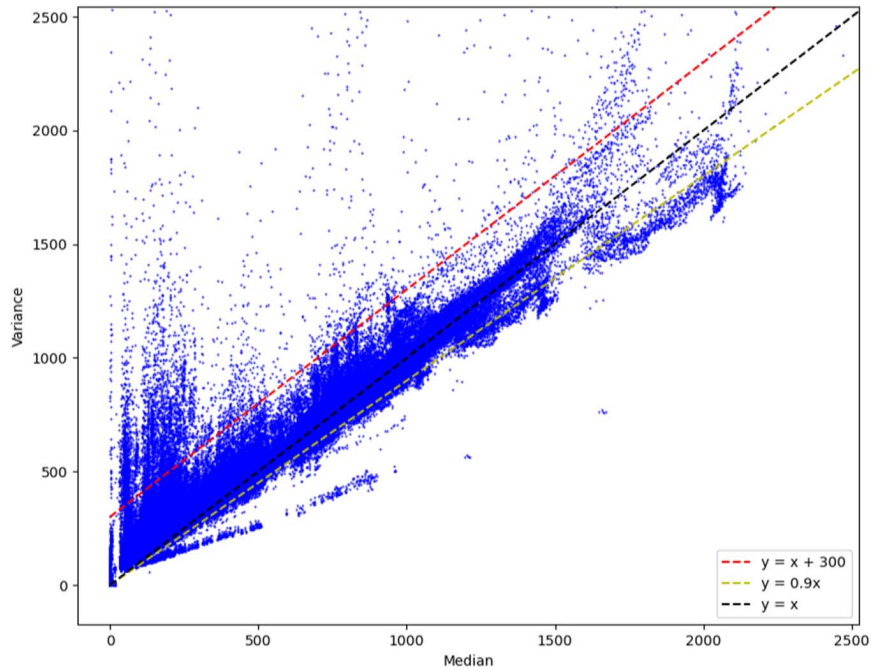

**Figure 3.** Median of background photon counts vs. variance of background photon counts for all chips in every image in the $g$, $r$, and $i$ bands. Shown in red is the $y = x + 300$ while the black line is $y = x$. The yellow line shows the $y = 0.9x$ line. If a chip is outside the strip which is defined by the red and yellow lines in over half the images, the chip is rejected. This condition was found to be effective in identifying severely defective chips.

To calculate photometric redshift, we use the magnitude brightness measured in the coadd of the five photometric bands. We use these as inputs to `EAZY` (G. B. Brammer et al. 2008) to determine photometric redshift. A few random galaxies across our field were selected and the photometric redshift was compared to the spectroscopic redshift obtained from the NASA Extragalactic Database (NED). These were generally found to be in good agreement.

### 2.5. PSF Correction

The PSF causes dilution in the ellipticity and shear signals and hence needs to be corrected. It has also been shown PSF variation across the image can masquerade as shear signal if not taken into account. PSF correction is done in two steps. First, we interpolate the PSF across the image; second, we then correct the shape and ellipticity dilution due to the PSF. In any astronomical image, the PSF can only be precisely determined at locations where stars are present. To determine the exact PSF shape at other locations, a variety of methods can be found in the literature. L. Van Waerbeke et al. (2002) used a second-order polynomial to model PSF variation across an image. Other elaborate methods have been proposed by H. Hoekstra (2004), J. Bergé et al. (2011), and C. Chang et al. (2012). We use an inverse distance weight interpolation, similar to one described by M. Gentile et al. (2012), because of its simplicity and performance. To find $\sigma_{xx}^2({\rm PSF})$, we use

$$\sigma_{xx}^2({\rm PSF}) = \sum_{i=1}^{n} \sigma_{xx,i}^2 w(i), \quad (6)$$

where $\sigma_{xx,i}^2$ is the measured $Q_{11}$ of the $i$th star obtained from the moment-matching method, and $n$ is the number of stars used. The weight $w(i)$ is defined as

$$w(i) = \frac{1/d_i}{\sum_i 1/d_i}, \quad (7)$$

where $d_i$ is distance of the $i$th star from the source. We found the number of stars selected for PSF correction in the range $n = 5$–25 does not affect our analysis significantly. We selected $n = 15$. A similar process is followed for interpolating $\sigma_{xy}^2({\rm PSF})$ and $\sigma_{yy}^2({\rm PSF})$. This completes PSF interpolation.

A significant portion (~35%) of these sources is measured to have a size smaller than the PSF. Clearly, these measurements are unphysical. Traditionally, sources smaller than the PSF have been rejected (D. Gruen et al. 2013; D. E. Applegate et al. 2014; J. McCleary et al. 2020). However, these sources contain a large amount of information which would be lost if we reject these sources. The size measurement being smaller than the PSF is simply due to statistical fluctuations. In other words, the sizes of both the PSF and sources have Poisson noise and other systematic effects, which give rise to size measurements that are smaller than the PSF. We introduce PSF correction using a novel Monte Carlo method. If we assume an elliptical Gaussian profile for both the source and PSF, the true $\sigma_{xx}^2$ of the source is

$$\sigma_{xx}^2(\text{true}) = \sigma_{xx}^2(\text{measured}) \pm \sqrt{\frac{S^4}{N}(1+K)} - \sigma_{xx}^2(\text{PSF}) \pm \epsilon(xx)_{\text{PSF}}, \quad (8)$$

where $\sigma_{xx}^2$(true) is the true value, $\sigma_{xx}^2$(measured) is the measured value, $\sigma_{xx}^2$(PSF) is the value for the PSF, and $\epsilon(xx)_{\text{PSF}}$ is the error in the PSF value. The expression $\sqrt{S^4(1+K)/N}$ is the Poisson error in $\sigma_{xx}^2$(measured) as shown by A. Dutta et al. (2024), where $S$ is the measured size, $N$ is the measured photon counts, and $K$ for the elliptical Gaussian case is $4\pi S^2 B/N$. $B$ is the background level. In order to find the error in the PSF, $\epsilon(xx)_{\text{PSF}}$, we use bright stars where the Poisson error is negligible. For the coadds, it was found the stars matched with the Gaia catalog are bright enough to neglect the Poisson error. This is also clear from Figure 5(a), where the red points show a constant width. This width corresponds to PSF variation due to turbulence and other systematics. We perform a $k=3\sigma$ clip for $\sigma_{xx}^2$(measured), $\sigma_{yy}^2$(measured), and $\sigma_{xy}^2$(measured) to reject any stars with significant blending. Stars brighter than magnitude 15.75 are rejected to avoid the brighter-fatter effect. This threshold is a factor of 2 lower than the threshold mentioned in the previous section. This is to ensure the star sample is well away from the limits at which the brighter-fatter effect becomes important. We also reject stars with bflag or vflag raised. The calculation of flags is presented in Section 2.6. The other flags are not considered since the stars matched with Gaia are fairly bright. Of the remaining stars, we find $P_{84} - P_{16}$ for $\sigma_{xx}^2$, where $P_i$ represents the $i$th percentile. This value corresponds to two standard deviations, i.e., one standard deviation from the median value on either side. Half of this value is defined as $\epsilon(xx, \text{Turbulence})_{\text{PSF}}$. This gives only the typical error across the frame due to turbulence. In Figures 6(a), (b), and (c), the histograms of stars used for PSF interpolation are shown. The red curve is a Gaussian centered on the median with a width equal to $\epsilon(xx, \text{Turbulence})_{\text{PSF}}$.

**Table 2**
List of SWarp Parameters Used on the Left and List of SExtractor Parameters Used on the Right

| Keyword | Value | Keyword | Value |
| --- | --- | --- | --- |
| WEIGHT_TYPE | MAP_WEIGHT | DETECT_MINAREA | 5 |
| COMBINE_TYPE | WEIGHTED | DETECT_THRESH | 0.8 |
| RESAMPLING_TYPE | LANCZOS3 | ANALYSIS_THRESH | 0.8 |
| SUBTRACT_BACK | Y | FILTER | Y |
| BACK_TYPE | AUTO | FILTER_NAME | gauss_11.0_13x13.conv |
| BACK_SIZE | 256 | DEBLEND_NTHRESH | 64 |
| BACK_FILTERSIZE | 2 | DEBLEND_MINCONT | 0.0001 |
| RESCALE_WEIGHTS | Y | CLEAN | Y |
| | | CLEAN_PARAM | 2.0 |
| | | WEIGHT_TYPE | MAP_WEIGHT |
| | | BACK_TYPE | AUTO |
| | | BACK_VALUE | 0.0 |
| | | BACK_SIZE | 128 |
| | | BACK_FILTERSIZE | 2 |

To the turbulence error we add the Poisson error of the stars used for PSF interpolation. Using the errors presented in A. Dutta et al. (2024) and standard error propagation, the PSF error due to Poisson noise is

$$\epsilon^2(xx, \text{Poisson})_{\text{PSF}} = \sum_{i=1}^{n} w^2(i) \left( \frac{S^4(i)}{N(i)} + \frac{4\pi BS^6(i)}{N^2(i)} \right)^2, \quad (9)$$

where $S(i)$ is the size of the $i$th star and $N(i)$ is the photon counts of the $i$th star. The Poisson and the turbulence components are added in quadrature to find $\epsilon(xx)_{\text{PSF}}$. It was found that the turbulence component always dominates in the coadds and is an order of magnitude larger than the Poisson component.

Now we have determined all the terms on the right-hand side of Equation (8). To find $\sigma^2$(true) we perform 30,000 iterations of the Monte Carlo Equation (8). In each iteration, the errors are randomly sampled from a Gaussian distribution with mean zero and standard deviation being the error value. This includes the Poisson error of both the source and the PSF and uncertainty in the PSF arising from turbulence, instrument, and optics. In each iteration sampling is done for $\sigma_{xx}^2$, $\sigma_{yy}^2$, and $\sigma_{xy}^2$. We only consider the cases where the following conditions are satisfied:

1. $\sigma_{xx}^2$(true) > 0;
2. $\sigma_{yy}^2$(true) > 0; and
3. $2\sigma_{xy}^2$(true) < $\sigma_{xx}^2$(true) + $\sigma_{yy}^2$(true).

The first two conditions ensure the size along the $x$- and $y$-axes individually are positive and hence the overall size is positive. The last condition ensures $|e_2| < 1$. We take median values of $\sigma_{xx}^2$, $\sigma_{yy}^2$, and $\sigma_{xy}^2$ in all the iterations that satisfy these conditions to find $\sigma_{xx}^2$(true), $\sigma_{yy}^2$(true), and $\sigma_{xy}^2$(true). We note the conditions listed above were found to be optimal. Several sets of conditions such as constraining overall size > 0 and constraining $|e_1| < 1$ were tried and rejected. Sources in which fewer than 50 samples satisfy the above conditions are considered to have failed PSF correction. Figures 7(a) and (b) show size histograms of stars and galaxies after PSF correction has been performed in the $i + r$ coadd.

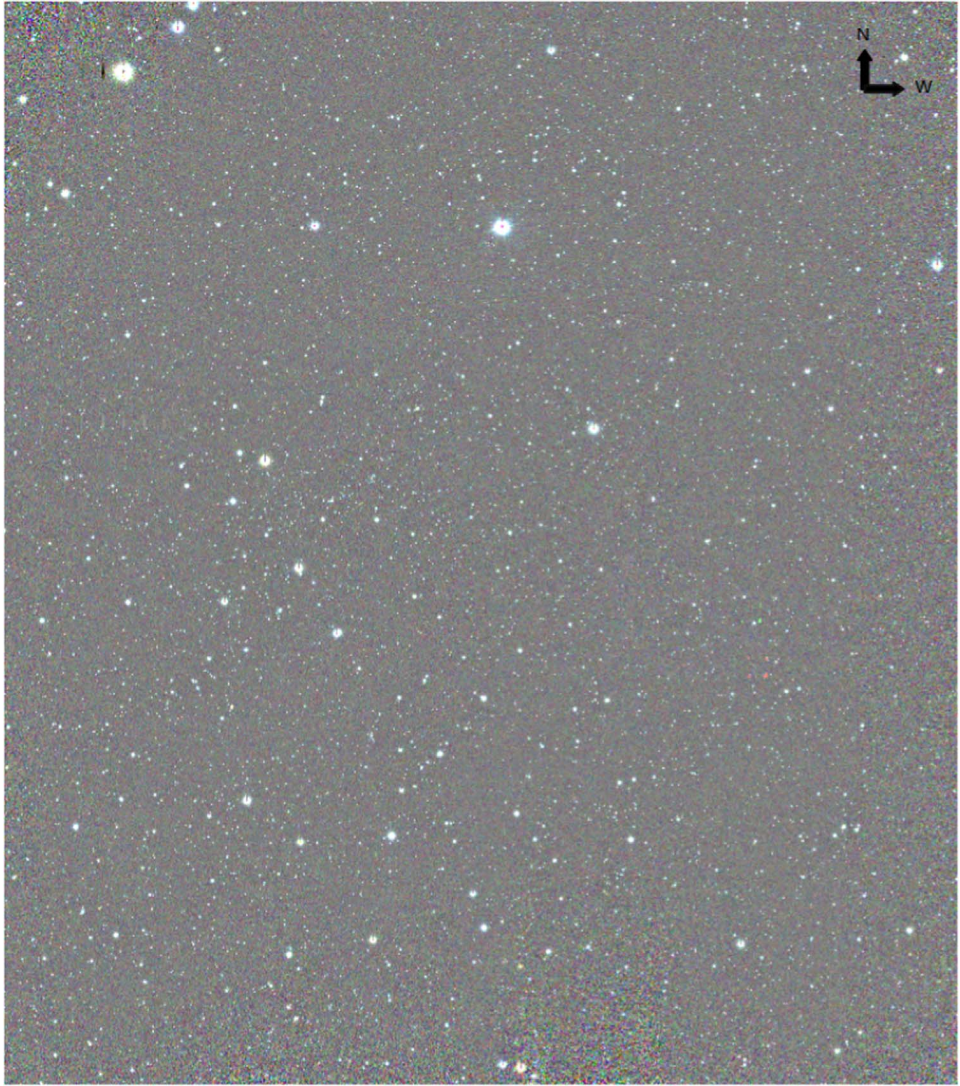

**Figure 4.** Red, green, and blue (RGB) color image of the full field containing the galaxy cluster A2390. We use the coadded images in the $z$, $i$, and $r$ bands, which correspond to the R, G, and B channels, respectively. The field of view is approximately $43' \times 49'$.

### 2.6. Flags

In this section, we describe the different flags used in this paper and the conditions that raise these flags. These flags are primarily used to determine which measurements are likely to be contaminated. For instance, sources that lie close to a chip edge or in chip gaps would need to be rejected. Sources that have a very small spatial separation or are blended would likely yield contaminated measurements as well. Below, we list all the flags used and the conditions that trigger the flag.

1. *Vertical flag* (`vflag`). Some brighter stars have a vertical array of zeros near the center. This can be seen in Figure 2(b). These stars are unsuitable for PSF determination. Let us assume the center of the source is at pixel $(x_c, y_c)$. The `vflag` is raised if any of the following conditions is satisfied:

   (a) Calculate the average pixel value in a horizontal $3 \times 1$ pixel strip two pixels above $y_c$. In other words, the mean value of pixels at $(x_c - 1, y_c + 2)$, $(x_c, y_c + 2)$, and $(x_c + 1, y_c + 2)$ is calculated. The mean value of pixels at $(x_c - 6, y_c + 2)$ and $(x_c + 6, y_c + 2)$ is also calculated. If the first quantity is less than either of the second quantities, then the flag is raised.
   (b) Similar calculation as before except we consider a strip three pixels above $y_c$. We find the mean value of pixels at $(x_c - 1, y_c + 3)$, $(x_c, y_c + 3)$, and $(x_c + 1, y_c + 3)$. The mean value of pixels at $(x_c - 7, y_c + 3)$ and $(x_c + 7, y_c + 3)$ is also calculated. If the former quantity is smaller than either of the latter quantities, the flag is raised.

2. *Bad flag* (`bflag`). As mentioned before some sources at the edges of CCDs or in chip gaps need to be rejected. We check if the cutout is at least 6 times the expected size of the source in that image. It was found that this is the minimum cutout size required to recover flux, shape, and ellipticity accurately. The expected size is calculated

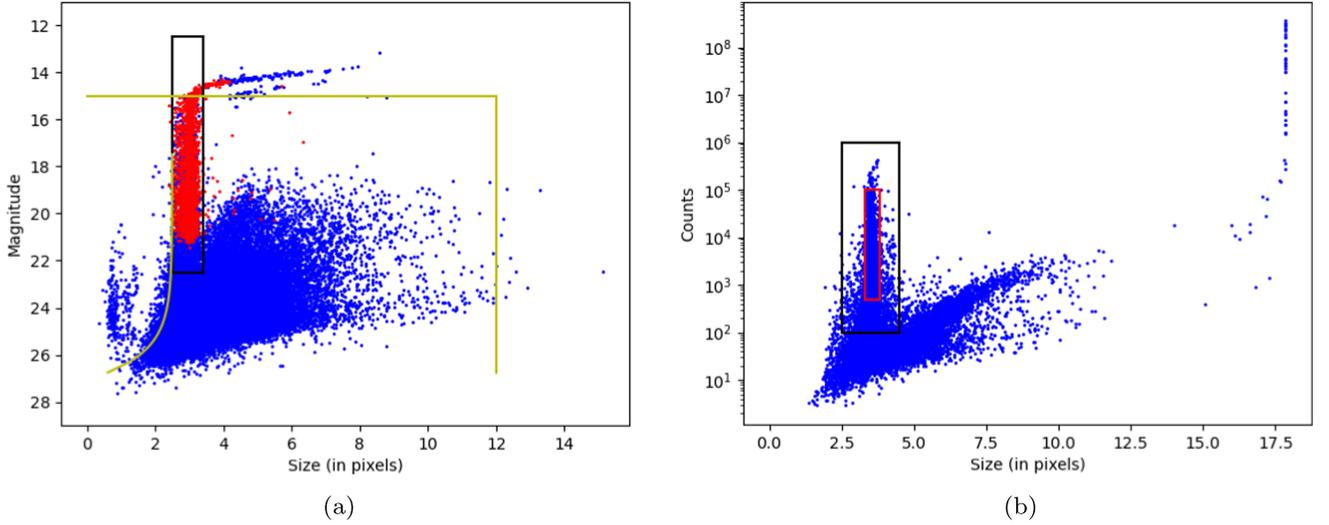

**Figure 5.** (a) Magnitude vs. size for sources detected and measured in the $i + r$ coadd. The points shown in red are stars matched with the Gaia EDR3 catalog. All sources inside the black box are rejected from the weak lensing analysis pipeline. So are the sources above the horizontal yellow line (15th magnitude) to account for the brighter-fatter effect. Sources to the right of the vertical yellow line are rejected to avoid blended sources and artifacts. Sources to the left of the curved yellow line are also rejected since these are three standard deviations smaller than the median PSF. (b) The photon counts vs. size graph for the $i + r$ coadded image simulated using PhoSim for the first case of $\gamma_1 = 0.1$. The points in the red box are used for PSF interpolation. The points inside the black box are not used for the shear analysis. The condition for the red box is $3.3 <$ size $< 3.8$ pixels and $500 <$ counts $< 10^5$. For the black box the conditions are $2.5 <$ size $< 4.5$ pixels and $10^2 <$ counts $< 10^6$.

using

$$\sigma^2_{\text{image}} = \sigma^2_{\text{coadd}}(\text{true}) + \sigma^2_{\text{image}}(\text{PSF}), \tag{10}$$

where $\sigma^2_{\text{coadd}}(\text{true})$ is obtained after using the Monte Carlo method described above to solve Equation (8) in the $i + r$ coadd. The interpolated values of the PSF in the individual image is $\sigma^2_{\text{image}}(\text{PSF})$. In this equation $\sigma^2$ can be substituted with $\sigma^2_{xx}$, $\sigma^2_{yy}$, or $\sigma^2_{xy}$. The expected size is

$$S = \sqrt{\sigma^2_{\text{image},xx} + \sigma^2_{\text{image},yy}}. \tag{11}$$

If the cutout size is larger than 6 times the expected size of the source, then we check for pixel values $\leqslant 0$ inside a square box of side 6 times the size. If found, this flag is raised, i.e., set to 1. If the initial cutout size is less than 6 times the size of the source being measured, any pixel value $\leqslant 0$ in the cutout would trigger this flag. This flag is exclusively used for cutouts made in individual frames.

3. *Background flag* (`bkgflag`). This flag is used in the $i + r$ coadd. We use the larger of $\sigma^2_{xx}$ or $\sigma^2_{yy}$ to find the size. Size is defined as $S = \sqrt{\sigma^2_{xx} + \sigma^2_{yy}}$. In this case, $\sigma^2_{xx}$ and $\sigma^2_{yy}$ are the values measured by the moment-matching method and not corrected for PSF. We replace the smaller of the two with the larger. We also find the average value of background standard deviation in the full image and call this global standard deviation $\sigma_g$. Next, assuming an elliptical Gaussian profile, we find the radius at which light from the source becomes equal to the background standard deviation. We denote this as $r$:

$$r = \sqrt{S^2 \log\left(\frac{2\pi S^2 \sigma_g}{N}\right)}, \tag{12}$$

where $N$ is the photon counts and $S$ is the size. We also define $r_1 = r/\sqrt{2}$. We now define eight regions. The first four regions are simply a $4 \times 4$ pixel square at the corners of a square with a size length $2r$. The next four regions are similar except now they are at the corners of a square of side length $2r_1$. We find the median value in each region. We also find the median value of the eight median values and call this the local median. If the local median is higher than the global median $+2$ times the global standard deviation, the flag is raised. We also count the number of regions whose median value is greater than the local median $+1.5$ times the local standard deviation or lower than the local median $-1.5$ times the local standard deviation. If more than one region exceeds this threshold, then the flag is raised as well.

4. *Bound flag* (`bndflag`). We calculate $r$ for all sources as described above. We cap the lower value of $r$ at 13 pixels. This number is decided since it is just larger than half the size of the minimum cutout of $26 \times 26$ pixels. For every source, the distance of the source centroid from the centroid of all other sources is calculated. If the centroid distance is lower than the sum of the $r$ for each pair considered, the flag is raised. This flag indicates if a significant amount of light from a source is present in the cutout of another source. Fewer than 2% of our sources were found to raise this flag. However, it was also found that this flag was not adequate in some cases because for most astronomical sources light falls off slower than a Gaussian.

### 2.7. Measurement in Individual Images

It has been suggested using measurements made in individual images over coadds could lead to better measurements (J. A. Tyson et al. 2008; M. J. Jee & J. A. Tyson 2011; R. Mandelbaum 2018). This is because coadding causes loss of information and averaging of systematic effects. In theory, measurements in individual images can be better corrected to take systematics into account. Measurements made in images with better seeing and lower background brightness can also be given more weight. Indeed, this idea has been used in some recent weak lensing studies such as L. Miller et al. (2013) and J. Zuntz et al. (2018). However, the challenge with such a scheme is that the sources will be extremely faint in each individual image. In all studies to date, measurements in individual images are performed only when the source has high enough SNR (∼10 or more). To get around this limitation, we use forced measurement (A. Dutta et al. 2024), which is capable of measuring sources with SNR ≪ 1. Giving the details of this algorithm is beyond the scope of this paper, but we describe it here briefly.

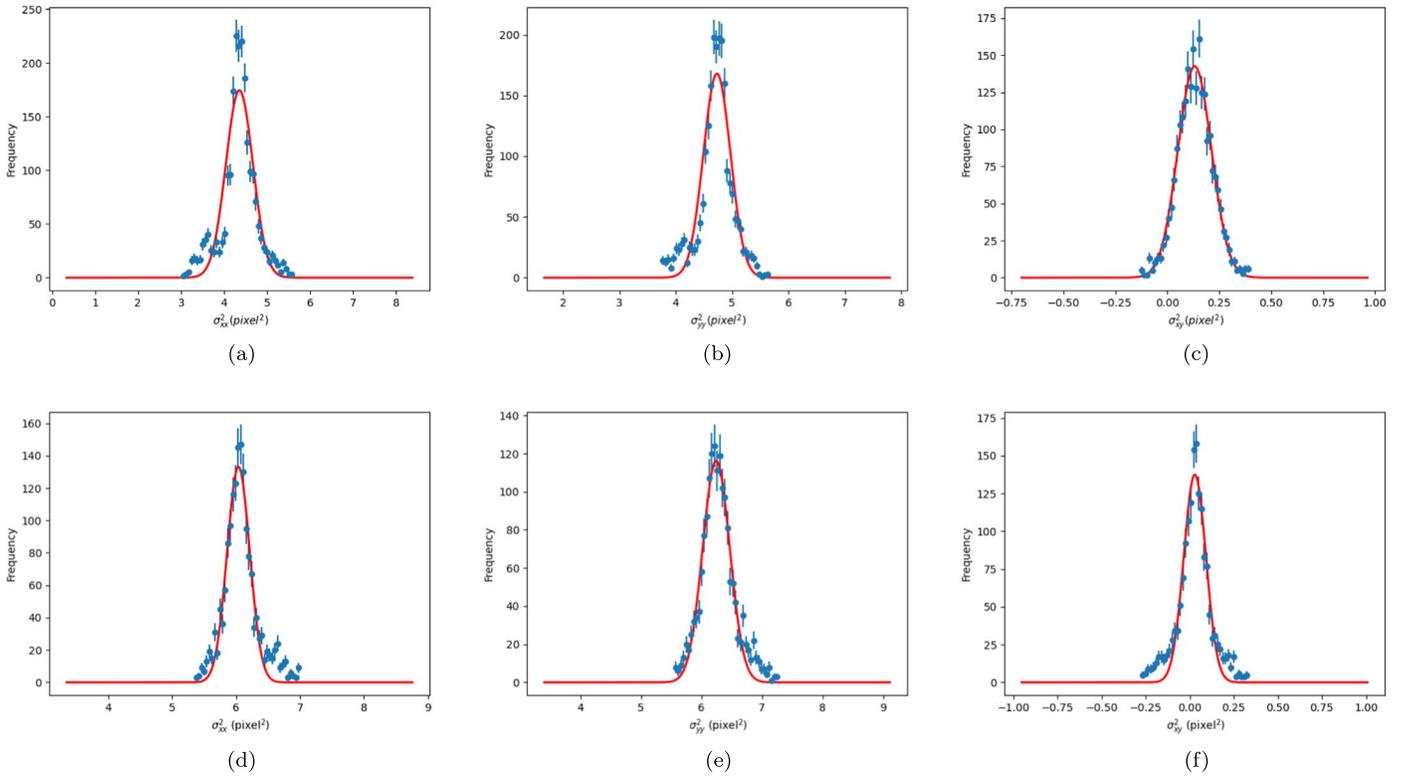

**Figure 6.** (a) The points show the $\sigma_{xx}^2$ of stars in the $i + r$ coadded image. The red curve shows the best-fit Gaussian obtained with a width half the value of $P_{84} - P_{16}$, where $P_i$ represents the $i$th percentile value. The fits are reasonably good though not perfect. (b) and (c) shows similar graphs for $\sigma_{yy}^2$ and $\sigma_{xy}^2$. The widths of the Gaussians are 0.37, 0.27, and 0.08, respectively. (d) The points show the $\sigma_{xx}^2$ of stars in the $i + r$ coadded image simulated using PhoSim for the first case, i.e., $\gamma_1 = 0.1$. The red curve shows the best-fit Gaussian for each case. (e) and (f) show similar graphs for $\sigma_{yy}^2$ and $\sigma_{xy}^2$. The widths of the Gaussians are 0.18, 0.22, and 0.06, respectively.

Forced measurement is a generalization of forced photometry (R. Lupton et al. 2001; C. Stoughton et al. 2002; D. Lang et al. 2016). Using the moment-matching algorithm described above, we can measure the shapes and positions of the detected sources in the coadd. Next, we can use this measurement as our initial guess and run the moment-matching algorithm for a single iteration, as opposed to convergence. The flux measurements obtained from this method are identical to forced photometry, and so it stands to reason that the centroid and shapes found could hold some information. This was found to be true and is the essence of forced measurement. First, the pixel values after background subtraction are truncated in a fashion that makes all pixel values positive. This step is crucial to ensure that measured $\sigma^2$ values are not negative. Then we use a method similar to forced photometry, where we use the expected shape and size of the source to perform a single iteration of the moment-matching algorithm. The values of flux shapes and sizes thus obtained are corrected to take into account the effect of truncation. The details of this algorithm have been described in A. Dutta et al. (2024).

We decided to perform measurements only in $g$-, $i$-, and $r$-band individual images since almost all sources detected in the coadd can be detected in these individual band coadds as well. To estimate the PSF in individual exposures, a $80 \times 80$ pixel cutout at the location of the stars is made and moment-matching measurement is done on these cutouts. To discard any stars with defects or blending, we perform a $k = 3\sigma$ clip of all the stars in $\sigma_{xx}^2$, $\sigma_{yy}^2$, and $\sigma_{xy}^2$. Bright stars with SNR $\geqslant 100$ and no flags raised are selected for PSF estimation. To avoid the brighter-fatter effect, any star having $\geqslant 10^6$ photon counts is rejected. We also reject any stars in chips where the weight has been set to zero during coaddition. This produces a very pure sample of stars with which to determine the PSF. Interpolation of PSF is done using Equation (6).

For the remaining sources, a square cutout 8 times the size of the source is created with the source at the center. To check whether the source is bright enough for the moment-matching algorithm to converge, the approximate SNR is calculated. It has been shown in A. Dutta et al. (2024) that when SNR $\geqslant 15$ convergence is guaranteed. SNR is defined as

$$\text{SNR} = \frac{N}{\sqrt{N + 4AB}}, \quad (13)$$

where $N$ is the photon counts from the source, $A$ is the area, and $B$ is the background. The background is taken from the header and is the global median sky background. An alternate approach to determine if the source is bright enough for moment matching to converge would be to run the measurement algorithm on the cutout to check for convergence. However, this approach suffers from the fact that it would be challenging to differentiate between the case where the source is bright enough for convergence and the algorithm converging on a nearby bright source or defect. Hence, this approach was not explored further. Calculating the SNR before actually measuring the source is not possible. Hence, the SNR calculation is done using the expected size, expected flux, and overall background of the image. The expected flux ($N_{\text{expected}}$) and expected size ($\sigma^2_{\text{image}}$) are calculated as

$$N_{\text{expected}} = N_{\text{coadd}} \times \left\langle \frac{N_{\text{image}}(\text{star})}{N_{\text{coadd}}(\text{star})} \right\rangle, \quad (14)$$

$$\sigma^2_{\text{image}} = \sigma^2_{\text{coadd}}(\text{true}) + \sigma^2_{\text{image}}(\text{PSF}), \quad (15)$$

where $\sigma^2_{\text{coadd}}(\text{true})$ is obtained after using the Monte Carlo method described above to solve Equation (8) in the $i + r$ coadd. The interpolated values of the PSF in the individual image is $\sigma^2_{\text{image}}(\text{PSF})$. In this equation $\sigma^2$ can be substituted with $\sigma^2_{xx}$, $\sigma^2_{yy}$, or $\sigma^2_{xy}$. The expected photon counts of a source in an image is $N_{\text{expected}}$, $N_{\text{coadd}}$ is the photon counts of the source in the corresponding band coadd, and $\langle N_{\text{image}}(\text{star})/N_{\text{coadd}}(\text{star}) \rangle$ is the ratio of photon counts of stars in the image to photon counts in the coadd. The angle brackets in the above equation indicate the median. To find the median value, only the stars that are used for PSF interpolation at the given location are utilized. This was done to minimize the effect of photometric variation across the image.

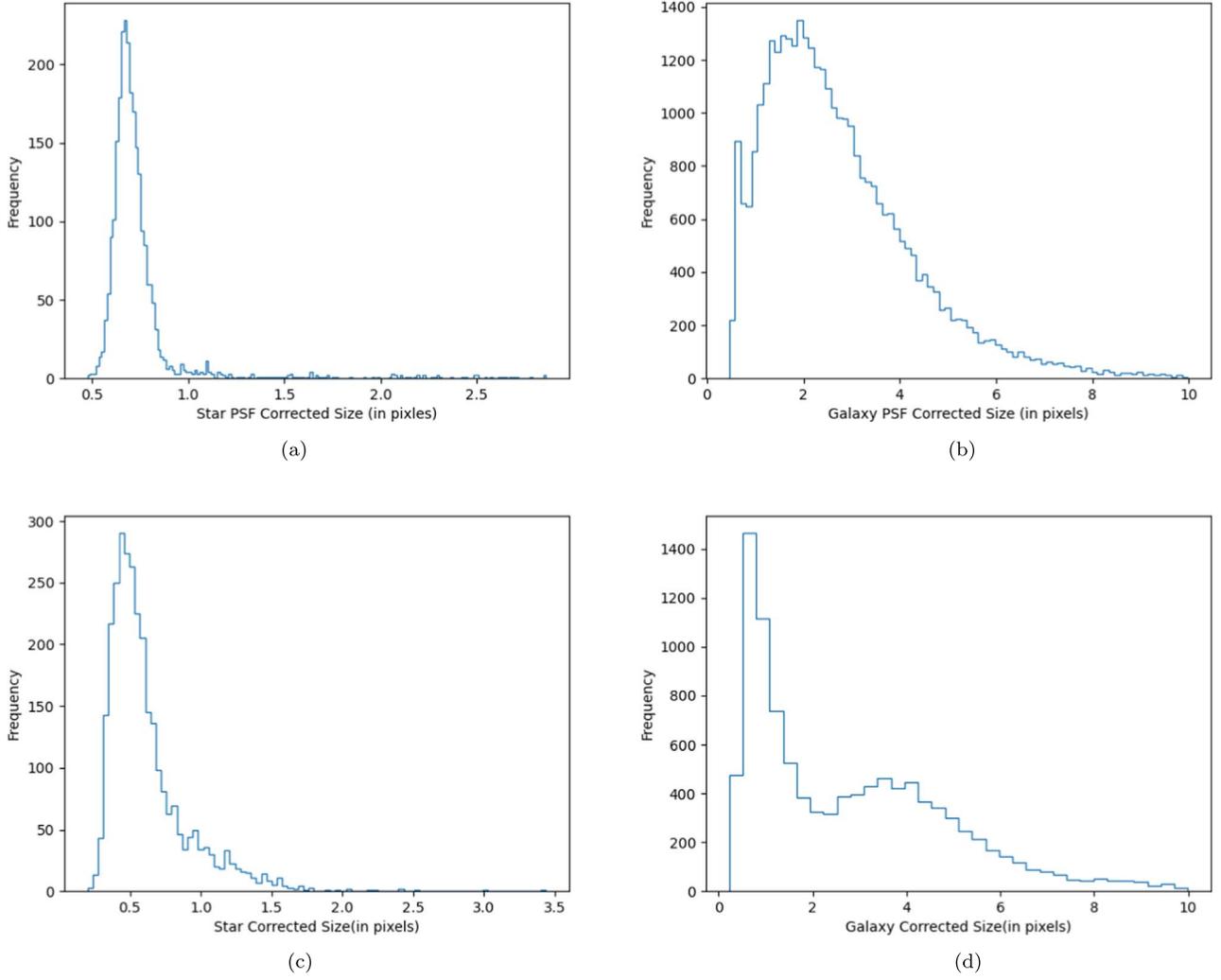

**Figure 7.** (a) and (b) Plot on the left shows a histogram of all stars after they have been passed through the Monte Carlo PSF correction method in the $i + r$ coadded image. Sigma clipping is necessary to reject the stars on the right tail, which are not suitable for calculating the PSF. The graph has been truncated at $x = 3$ pixels. A similar graph for galaxies is shown on the right. This graph has been truncated at $x = 10$ pixels. (c) and (d) Plot on the left shows a histogram of all stars after they have been passed through the Monte Carlo PSF correction method in the $i + r$ coadded images simulated using PhoSim with input $\gamma_1 = 0.1$. The plot is truncated at $x = 3.5$ pixels. Not all stars shown in the histogram are used for PSF correction. Sigma clipping is required to reject the stars on the right tail, which are not suitable for calculating the PSF. A similar graph for galaxies is shown on the right. This plot is truncated at $x = 10$ pixels.

If SNR is lower than 15, we use forced measurement (A. Dutta et al. 2024). Forced measurement requires a reasonably good starting guess. We use $N_{\text{expected}}$ and $\sigma^2_{\text{image}}$ from the above equations as guess values for forced measurement. After forced measurement has been performed, we similarly perform Monte Carlo PSF correction as before. A slightly modified Monte Carlo equation shown in Equation (17) is used to take into account that the errors in the flux, shape, and size obtained from forced measurement are significantly tighter than Poisson errors. This arises from the fact that forced measurement needs a reasonable guess to start with and only a single iteration is performed. The errors in the measured parameters are calculated as

$$\epsilon_{\sigma_{xx}^2} = p_{\sigma_{xx}^2}(N, S, B)\sqrt{\frac{S^4}{N} + 4\pi S^2 B \frac{S^4}{N^2}}, \quad (16)$$

where $N$ is the photon count, i.e., $N_{\text{expected}}$, $S$ is the size, and $B$ is the background value; $p_{\sigma_{xx}^2}$ is a function of $N_{\text{expected}}$, $S$, and $B$. We use the piecewise linear function of $p_{\sigma_{xx}^2}$ from A. Dutta et al. (2024). All successful measurements are passed through the Monte Carlo PSF correction scheme similar to Equation (8). For individual exposures the PSF correction equation for $\sigma_{xx}^2$ is

$$\sigma_{xx}^2(\text{true}) = \sigma_{xx}^2(\text{measured}) \pm p_{\sigma_{xx}^2}\sqrt{\frac{S^4}{N}(1+K)}$$
$$- \sigma_{xx}^2(\text{PSF}) \pm \epsilon(xx)_{\text{PSF}}, \quad (17)$$

where $\sigma_{xx}^2(\text{measured})$ is the measured values in individual exposures by forced measurement. In cases where the moment-matching method successfully converges, $p_{\sigma_{xx}^2}$ is set to 1. The error in the PSF is $\epsilon(xx)_{\text{PSF}}$. A similar equation is used to determine $\sigma_{yy}^2(\text{true})$ and $\sigma_{xy}^2(\text{true})$. We perform 30,000 iterations and use the same condition used in the coadd PSF correction to select which iterations as considered valid. Sources for which fewer than 50 samples satisfy the above conditions are considered to have failed PSF correction.

### 2.8. Combining Data from Individual Images to Get Shear

The individual frame measurements of each source need to be combined into a single shape and ellipticity measurement. To reject contaminated measurements, we apply some sensible cuts. We reject all measurements where any flag has been raised either in the $i+r$ coadd or individual frames. Sources where the size is over a factor of 2 larger than the $i+r$ coadd size after PSF correction are rejected. If this condition fails, it is a strong indication that the source was not measured accurately. We reject measurements where the flux value is negative or not a number (NaN). This is because negative or NaN values do not make physical sense. Cases where PSF correction failed in single frames and cases where forced measurement produces NaNs are rejected as well. Finally, to ensure defects and noise do not affect our measurements, we calculate the sigma-clipped median and standard deviation of individual frame flux measurements in a filter. Any measurement where the flux is three standard deviations away from the median is rejected. This is to ensure defects like the "stripes" shown in Figure 2(a) do not influence the measurement of faint sources. We also reject images with $\sigma_{xx}^2(\text{PSF})$ or $\sigma_{yy}^2(\text{PSF})$ greater than 30 pixel$^2$, which corresponds to a size of approximately 7.7 pixels, i.e., a seeing FWHM of 2″6. Weak lensing analysis requires very stringent PSF cuts, and images with seeing larger than 1″ have traditionally been discarded (N. Okabe et al. 2010; D. Gruen et al. 2013). Also, at approximately 4″ seeing point sources start to form donut shapes, indicating the telescope is extremely out of focus. We also reject images where the sigma-clipped standard deviation in stellar $\sigma_{xx}^2(\text{PSF})$ or $\sigma_{yy}^2(\text{PSF})$ is greater than 3 pixels. This corresponds to an unusually large variation of PSF across the field and is indicative of issues with the image. These conditions are not mutually exclusive.

The images with better seeing conditions and lower background brightness should be given more weight. Put differently, images which have the lowest Poisson error should be given maximum weight. We use inverse of the error in the second moments as weight when combining the $\sigma_{xx}^2$, $\sigma_{yy}^2$, and $\sigma_{xy}^2$ of individual measurements. This was found to be optimal:

$$\sigma_{xx}^2 = \sum_i \sigma_{xx,i}^2 W(i), \quad (18)$$

where $\sigma_{xx,i}^2$ is the PSF-corrected measurement in the $i$th frame and the weight of the $i$th image $W(i)$ is

$$W(i) = \frac{\epsilon_{\sigma_{xx,i}^2(\text{true})}^{-1}}{\sum_{i=1}\epsilon_{\sigma_{xx,i}^2(\text{true})}^{-1}}, \quad (19)$$

where $\epsilon_{\sigma_{xx}^2(\text{true})}$ is the error in $\sigma_{xx}^2(\text{true})$. This is equivalent to $\epsilon_{\sigma_{xx}^2}$ and $\epsilon(xx)_{\text{PSF}}$ added in quadrature:

$$\epsilon_{\sigma_{xx,i}^2(\text{true})} = \sqrt{\epsilon_{\sigma_{xx,i}^2}^2 + \epsilon^2(xx)_{\text{PSF},i}}. \quad (20)$$

A similar equation for $\sigma_{yy}^2$ and $\sigma_{xy}^2$ is used. After combining the individual measurements, the error in the final $\sigma_{xx}^2$ is

$$\epsilon_{\sigma_{xx}^2(\text{true})} = \sqrt{\sum_{i=1} \epsilon_{\sigma_{xx,i}^2(\text{true})}^2 W^2(i)}. \quad (21)$$

Ellipticities $e_1$ and $e_2$ are calculated as

$$e_1 = \frac{\sigma_{xx}^2 - \sigma_{yy}^2}{\sigma_{xx}^2 + \sigma_{yy}^2}, \quad (22)$$

$$e_2 = \frac{2\sigma_{xy}^2}{\sigma_{xx}^2 + \sigma_{yy}^2}. \quad (23)$$

The error in the final ellipticity measurement can be calculated using standard error propagation equations. This comes out to

$$\epsilon_{e1}^2 = 4\epsilon_{\sigma_{xx}^2(\text{true})}^2 \frac{\sigma_{xx}^4(\text{true}) + \sigma_{yy}^4(\text{true})}{(\sigma_{xx}^2(\text{true}) + \sigma_{yy}^2(\text{true}))^4}, \quad (24)$$

$$\epsilon_{e2}^2 = 8\epsilon_{\sigma_{xx}^2(\text{true})}^2 \frac{\sigma_{xy}^4(\text{true})}{(\sigma_{xx}^2(\text{true}) + \sigma_{yy}^2(\text{true}))^4}$$
$$+ 4\epsilon_{\sigma_{xy}^2(\text{true})}^2 \frac{1}{(\sigma_{xx}^2(\text{true}) + \sigma_{yy}^2(\text{true}))^2}. \quad (25)$$

In the above derivation we have assumed $\epsilon_{\sigma_{xx}^2(\text{true})} = \epsilon_{\sigma_{yy}^2(\text{true})}$. This is approximately true as long as the PSF and source are not extremely elliptical. Considering only a small fraction of the sources show extremely high ellipticity, this error equation is reasonably close to the true Poisson value.

On comparing the size measured from the $i+r$ coadded image to the the combined measurement from individual images, we see a significant reduction in size for a large fraction of sources. This is shown in Figure 8(a). A smaller size increases the shear signal. In Figure 8(b), we show the same graph for simulated sources. The simulation is described in Section 3.

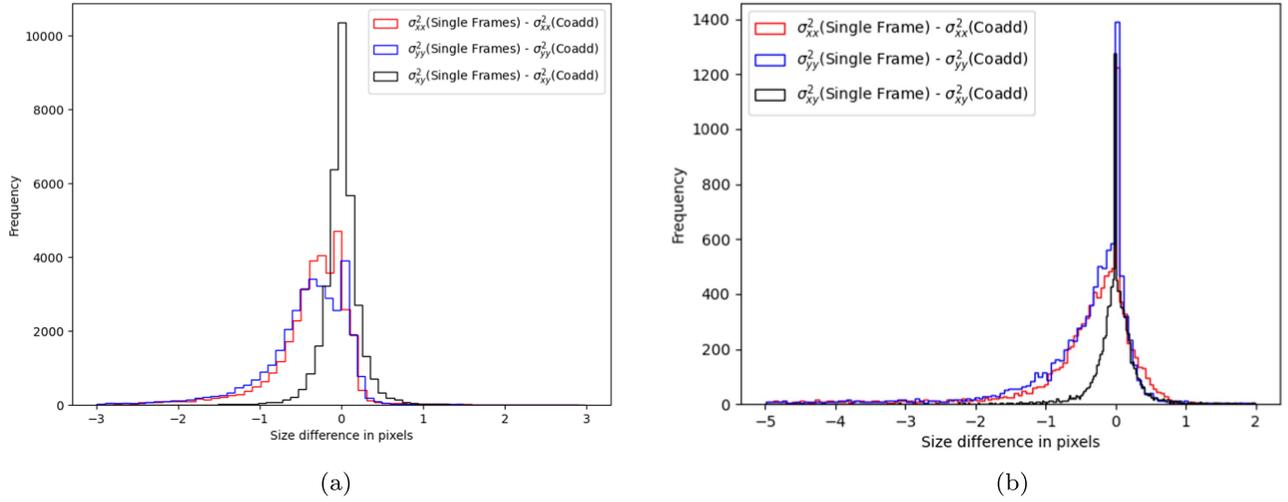

**Figure 8.** (a) Histograms for the difference in shape when sources are measured from the $i + r$ coadd vs. when they are measured in individual images of $i$- and $r$-band images. The individual image measurements are combined using inverse error as weight. The reduction in size by weighting individual frames optimally is clear from this plot. This leads to enhanced shear signal and hence shear recovery. (b) Histograms for the difference in shape when sources are measured from the weighted $i + r$ coadd vs. when they are measured in individual simulated images of $i$- and $r$-band images produced using PhoSim for the first case, i.e., $\gamma_1 = 0.1$. The individual image measurements are combined using inverse error as weight. The reduction in size by weighting individual frames optimally is clear from this plot. This leads to enhanced shear signal and hence shear recovery.

To recover shear from ellipticities, we adopt the method proposed in the GREAT3 challenge (R. Mandelbaum et al. 2014):

$$g_t \approx \gamma_t = \frac{e_t}{2(1 - \langle e^2 \rangle)}, \quad (26)$$

where $\gamma_t$ is the tangential ellipticity, $e_t$ is the tangential ellipticity of a source with respect to some given central point, and $\langle e^2 \rangle$ is the average value of ellipticity squared. Reduced tangential shear is denoted by $g_t$. This equation was slightly modified to take into account the weight of each source:

$$g_t \approx \gamma_t = \frac{\sum_{i=1} e_{t,i} w_{a,i}}{2(1 - \langle e^2 \rangle)}, \quad (27)$$

where $e_{t,i}$ is the tangential component of ellipticity of the $i$th source about a given point and the weight $w_{a,i}$, the weight of the $i$th source, is defined as

$$w_{a,i} = \frac{\epsilon_{e,i}^{-2}}{\sum_{i=1} \epsilon_{e,i}^{-2}}, \quad (28)$$

where $\epsilon_e$ is the error in ellipticity and

$$\epsilon_e^2 = \epsilon_{e1}^2 \frac{e_1^2}{e_1^2 + e_2^2} + \epsilon_{e2}^2 \frac{e_2^2}{e_1^2 + e_2^2}, \quad (29)$$

where $\epsilon_{e1}^2$ is the error in $e_1$, and $\epsilon_{e2}^2$ is the error in $e_2$. The tangential ellipticity, $e_t$, defined about a point can be written as

$$e_t = -e_1 \cos(2\phi) - e_2 \sin(2\phi), \quad (30)$$

where $\phi$ is the angle a line joining the point and the source makes, and $e_1$ and $e_2$ are the ellipticity components. The cross component of ellipticity is similarly written as

$$e_c = -e_1 \sin(2\phi) + e_2 \cos(2\phi). \quad (31)$$

We also define

$$\langle e^2 \rangle = \sum_{i=1} e_i^2 w_{b,i}. \quad (32)$$

The weight $w_{b,i}$ is defined as

$$w_{b,i} = \frac{\epsilon_{e,i}^{-1}}{\sum_{i=1} \epsilon_{e,i}^{-1}}. \quad (33)$$

Using error propagation, the error in shear $\epsilon_\gamma^2$ when using Equation (27) is

$$\epsilon_\gamma^2 = \gamma^2 \left[ \frac{\sum_i w_{a,i}^2 \epsilon_{e,i}^2}{\left(\sum_i w_{a,i} e_{\tan,i}\right)^2} + \frac{2 \sum_i w_{b,i}^2 \epsilon_{e,i}^2}{1 - \left(\frac{\sum_i e_i^2 w_{b,i}}{\sum_i w_{b,i}}\right)^2} \right], \quad (34)$$

where the summation is performed over all sources being considered, $\gamma$ is the shear obtained, $\epsilon_{e,i}$ is the error in the ellipticity of the $i$th sources, $e_{\tan,i}$ is the tangential ellipticity component of the $i$th sources, and $e_i^2$ is the ellipticity of the $i$th sources. The uncertainty in shear given by Equation (34) is entirely due to Poisson noise. However, when considering real-world data the error in shear is significantly worse due to systematic errors in telescopes, turbulence of the atmosphere, imperfections in CCDs, and a multitude of other effects. Single-frame measurements will be affected the most since most of the sources are extremely faint in the individual images of a coadd. Hence, simulations are performed to understand these additional factors better.

### 3. Validation

#### 3.1. Simulation and Coadd Measurements

In order to test the ability of the pipeline described in Section 2 to recover shear accurately, we test it on simulated images produced by PhoSim (J. R. Peterson et al. 2015, 2019, 2020, 2024). PhoSim is a ray-tracing software that can simulate most of the known physics and hence produce extremely realistic images. We use the WIYN-ODI instrument, which has been validated with real data from the telescope, and simulate 30 images in the $i$ and $r$ filters with 60 s exposure time.

The seeing and air mass were chosen at random from real $i$- and $r$-band images. We chose to simulate sources fainter than 17th magnitude to prevent saturation. We perform two sets of simulations in two different areas of the sky. This ensures we have different ensemble properties for the two cases. In the first case, the galaxies in the PhoSim catalog were sheared by $\gamma_1 = 0.1$, and in the second case by $\gamma_1 = 0.05$. We pass the images through the exact coadd and detection method described above. The final coadd size is cropped to $10{,}000 \times 10{,}000$ pixels or $18\rlap{.}'3 \times 18\rlap{.}'3$. Source density in both cases is approximately 55 sources arcmin$^{-2}$. While on one hand, such a high number density from ground-based images is unlikely, on the other hand, it will help to test the performance in a crowded environment where blending issues pose a significant challenge for weak lensing analysis (J. Hartlap et al. 2011; W. A. Dawson et al. 2015).

For each image simulated, a $8 \times 8$ star grid is also simulated with the exact same seed, which ensures all random parameters used in PhoSim are identical. This is done for two reasons. One, the clouds in each individual image are slightly different and hence the ZP will be slightly different. By simulating a star grid with stars of known magnitude we are able to calculate ZP accurately. Second, it helps us get an initial estimate of seeing and background. Our pipeline needs ZP, seeing, and background variance to create the weights for the coadd. The magnitude brightness of stars in this grid ranges from 16 to 23.

The simulated images were passed through the same pipeline as above. The PSF size of the $i + r$ coadded image is 3.5 pixels and the errors in $\sigma_{xx}^2$, $\sigma_{yy}^2$, and $\sigma_{xy}^2$ are 0.18, 0.22, and 0.06, respectively. The size and errors are very similar for both sets of simulations since the same values of seeing, air mass, and seed were used for both. While the average size of the PSF is comparable to the real data, the errors are significantly smaller. The flux versus size graph is shown in Figure 5(b). Points inside the red square are used for PSF estimation, while the sources inside the black box are not used for weak lensing analysis to avoid shear dilution due to stars. The conditions for the red box are $3.3 <$ size $< 3.8$ pixels and $500 <$ counts $< 10^5$. For the black box, the conditions are $2.5 <$ size $< 4.5$ pixels and $10^2 <$ counts $< 10^6$. These limits were obtained after careful visual examination. We also reject any sources larger than 12 pixels, which are likely due to severe cases of blending or sources brighter than $10^5$ photon counts where the brighter-fatter effect becomes important. This is clearly seen when considering the sources just above the red rectangle in Figure 5(b). In total, the fraction of sources rejected by these cuts is roughly one-third, most of them being stars enclosed by the black box in Figure 5(b). This is in agreement with the PhoSim input catalog in which approximately one-third of all sources are stars. In Figures 7(c) and (d), we show histograms of star and galaxy size after PSF correction. The histogram of galaxy size shows a significant bump at 0.5 pixels, which implies contamination from stars. This could potentially explain the slight underestimation of shear in Figure 9. The histogram of star size along with the best-fit Gaussian in red is shown in Figures 6(d), (e), and (f).

It was found that a few percent of the sources with extremely small ellipticity errors, primarily large and bright sources, bias the shear measurement when using this scheme as described in the previous section. Capping ellipticity error to one-third of the median ellipticity error values works well to recover shear and hence was adopted. Using this simple modification we are accurately able to recover shear, as shown in Figure 9. In this figure, we plot the shear measured as we start from a central circular region and increase the radius gradually to include more sources. The $x$-axis shows the radius of the circular region. On the $y$-axis we plot the recovered shear. We find the shear recovered from individual image measurements is more accurate than using shape measurements made in the coadd. In the second case, there seems to be a slight bias of $\sim 0.005$. It was also found $\epsilon_e^2$, i.e., the Poisson component of ellipticity error does not accurately predict error bars. This is expected since the simulations accurately take into effect most known physics while our error bars only take into account the Poisson statistics. We found 0.005 to be a more suitable error bar. This error bar also allows us to ignore the bias since it is comparable to the error bar. This error level is approximately 4 times the Poisson error calculated from the coadd and 10 times the error calculated from combined single-frame measurements. These factors were adopted and applied to the Poisson error obtained from WIYN-ODI.

## 4. Weak Lensing Analysis

### 4.1. Aperture Mass Maps

A common method of detecting mass concentrations using weak lensing is aperture mass statistics (N. Kaiser 1995; P. Schneider 1996; D. Gruen et al. 2013). This measures the average tangential orientation of background galaxies around some point convolved with a weight. Aperture mass $M_{\rm ap}$ and the associated error $\sigma_{M_{\rm ap}}$ are defined as

$$M_{\rm ap} = \sum_i w(|\theta| - |\theta_i|) w_p(i) g_{i,t}, \quad (35)$$

$$\sigma_{M_{\rm ap}} = \sqrt{\sum_i w^2(|\theta| - |\theta_i|) w_p(i)}, \quad (36)$$

where $g_{i,t}$ is the tangential alignment of the $i$th galaxy and $w_p(i)$ is the weight function derived from Poisson error in ellipticity, and the overall weight function is denoted by $w(|\theta|)$. $\theta$ represents the generalized coordinates. Specifically, the weight due to inverse Poisson error in ellipticity is

$$w_p(i) = \frac{1}{\epsilon_e^2(i)}. \quad (37)$$

Several different weighting schemes can be found in the literature (P. Schneider 1996; M. Schirmer et al. 2004; D. Gruen et al. 2013). We use Gaussian weight (D. Gruen et al. 2013) for simplicity:

$$w(|\theta|) = \begin{cases} \exp(-|\theta|^2/2\sigma_w^2) & \text{when } |\theta| < 3\sigma_w \\ 0 & \text{when } |\theta| > 3\sigma_w \text{ or } |\theta| < 50 \text{ pixels} \end{cases}, \quad (38)$$

where $\sigma_w$ is the width of the weight function. The size of $\sigma_w$ is usually a few arcminutes. A smaller value brings out details at a finer spatial scale but at the cost of more noise. This is because, for a smaller width, averaging of shear is done in a smaller area using a small number of background galaxies. A larger value of width washes out details of mass distribution but has a lower noise.

For Poisson error in ellipticity $\epsilon_e$, we apply the error cap at one-third the median error value in order to prevent the measurements from being dominated by bright sources with extremely low errors. The significance maps are then defined as $M_{ap}/\sigma_{M_{ap}}$. $M_{ap}$ is also called the aperture mass maps or E-mode maps. To make the B-mode maps the tangential component of shear is replaced by the cross component. In other words, we simply replace $g_t$ with $g_c$ in Equation (35). This is defined as

$$g_c \approx \gamma_c = \frac{e_c}{2(1 - \langle e^2 \rangle)}. \qquad (39)$$

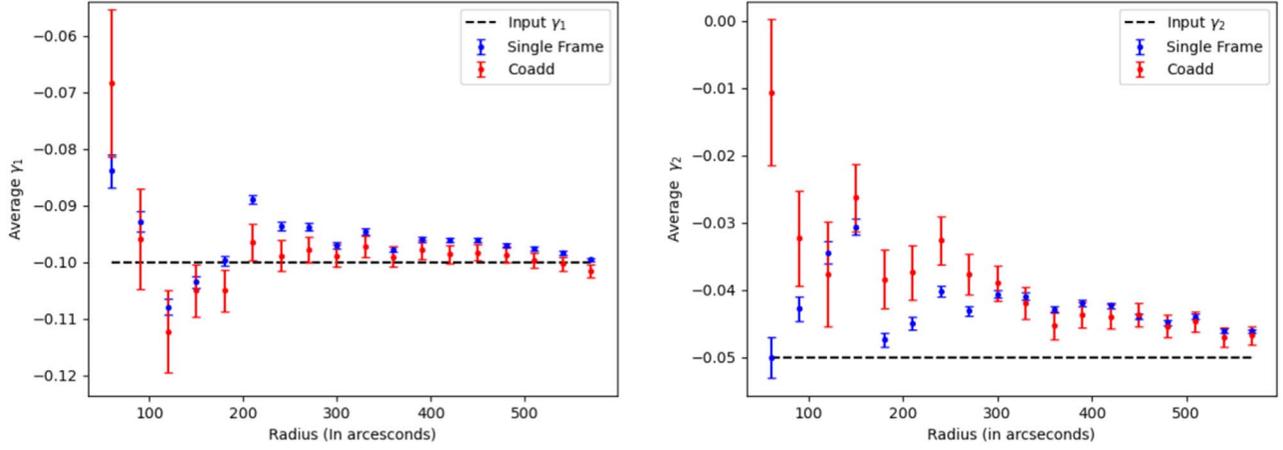

**Figure 9.** These plots show the ability of our pipeline to accurately recover shear at various levels in the weak lensing regime. To make these plots, we give the PhoSim simulations an input shear of $\gamma_1 = 0.1$ in the first case and $\gamma_2 = 0.05$ for the second case. It is clear both coadd and individual frame measurements accurately recover shear to within 0.005 of the true shear. In the second case, there is a slight bias of approximately 0.005. The error bars shown are just the Poisson component. Clearly, single frames have smaller error bars as expected. However, the error bars fall short because of non-Poisson sources of error. Hence, we elect to use error bars of 0.005 for shear measured using individual image measurements, which is approximately 10 times the Poisson component.

A major issue in making the aperture mass maps is that source density varies significantly over a large field. In our case, the source density varies by a factor of 2 in the $40' \times 43'$ field of view. This produces mass maps that have different error levels in different regions of the image. The problem gets particularly worse at the edges. Hence, we decide to use an adaptive weighting scheme. Simply put, we change $\sigma_w$ depending on the source density. The width is smaller in regions of high source density and, conversely, the width is larger when the source density is smaller. The width is changed to get $\sigma_{M_{ap}}$ varying by at most 10% throughout the field, except at the edges. However, we limit the maximum width to 3000 pixels and the minimum width to 500 pixels. The maximum value of width is reached only at the edges.

In Figure 10, we show the comparison of the aperture mass map with the light density maps in the central part of the cluster. To make the light density maps we select all galaxies with photometric redshift between 0.15 and 0.55 where the reliability parameter of EAZY is greater than 0.8. We also reject galaxies where the source was measured in fewer than three bands. This range includes the galaxy cluster A2390, which has been spectroscopically confirmed at $z = 0.23$ (R. G. Abraham et al. 1996). We reject any galaxy with size greater than 10 pixels. Then we divide our images into grids of $50 \times 50$ pixels. For each grid position we calculate the light from the galaxies in a $5'.5$ radius and weigh them by a Gaussian of width $0'.73$. The adaptive E-mode maps are made by fixing the value of $\sigma_{M_{ap}}$ within the range $3.5 \times 10^{-4} \pm 1.75 \times 10^{-5}$ and considering the galaxies in the redshift range $z = 0.4$–2.0. We do not consider galaxies with $z > 2.0$ since in this range a large fraction of the sources represent catastrophic failure of the photo-$z$ code. Once again, we only consider galaxies where the reliability score of photometric redshift is greater than 0.8 and flux information is available in at least three filters. We also reject sources larger than 10 pixels in size or sources where the bkgflag is raised. These conditions are not mutually exclusive and there is significant overlap. Changing the conditions slightly does not significantly affect our results. The main peaks of the E-mode maps line up extremely well with the galaxy density maps, which in turn line up with the central cD galaxy and X-ray maps (S. Allen et al. 2001). In Figure 11(a), we show the contours of the aperture mass map overlaid on the light density map of galaxies in the redshift range $z = 0.15$–0.55. The symbols show the location of galaxy clusters and groups obtained from the NED. Figure 12 shows the light density map of galaxies in the redshift range $z = 0.35$–0.75 with contours of aperture mass map overlaid on it. We discuss these maps in detail in Section 6.

### 4.2. Mass Maps

Aperture mass maps are useful for locating peaks and the general structure of a mass distribution. However, one of the key objectives of weak lensing studies is the measurement of mass. We follow the method of N. Kaiser & G. Squires (1993), which was later generalized by C. Seitz & P. Schneider (1995) to estimate the mass. $\kappa(\theta)$ is given as

$$\kappa(\theta) \approx \frac{1}{2\pi\bar{n}} \sum_{n=1}^{N} \frac{W(\theta - \theta_n, s)\gamma_{n,t}}{\theta_n^2}, \qquad (40)$$

where $\gamma_{n,t}$ is the tangential shear of the $n$th source, $\bar{n}$ is the density of sources, and

$$W(x, s) = 1 - \left(1 + \frac{x^2}{2s^2}\right)\exp\left(-\frac{x^2}{2s^2}\right), \qquad (41)$$

where $x$ is the distance of the source from the point about which we calculate $\kappa(\theta)$ and $s$ is the window width. The weight function $W(x)$ in effect suppresses noise at a small radius. If a very small window function is chosen, then the image becomes extremely noisy. On the other hand, if a large value of $s$ is chosen, then the details of the mass distribution are washed out. It was found that s = 200 pixels or 22″ produces an optimal result. To produce the $\kappa$ map the exact same procedure used to make E-mode maps is followed, except we now no longer use adaptive values. We also only consider galaxies in a radius of $7\farcm33$ for any given point. The map of $\kappa(\theta)$ for the entire field is shown in Figure 11(b). The error level in each pixel was determined using error propagation and multiplying the resulting Poisson error with a factor of 10 to take into account systematic/non-Poisson sources of error, as determined from PhoSim simulations. To convert from $\kappa(\theta)$ to enclosed mass we use

$$\kappa(\theta) = \frac{\Sigma}{\Sigma_{\text{crit}}}, \quad (42)$$

where $\Sigma$ is the mass enclosed and

$$\Sigma_{\text{crit}} = \frac{c^2}{4\pi G} \frac{D_s}{D_l D_{ls}}, \quad (43)$$

where $D_s$ is the angular distance to the source, $D_l$ is the angular distance to the lens, and $D_{ls}$ is the angular distance measured by an observer at the location of the lens to the source. This is not the same as simply subtracting $D_s$ and $D_l$ and depends on the curvature of the Universe. It should be noted that $\Sigma_{crit}$ is defined for a single source. However, in weak lensing analysis, one typically has thousands of background sources at various redshifts for a given lens system. In our case, the lens system is A2390. Hence, an average value of $\frac{D_s}{D_{ls}}$ is determined to find $\Sigma_{crit}$. The average value is defined as (N. Okabe et al. 2010)

$$\left\langle \frac{D_s}{D_{ls}} \right\rangle = \int \frac{D_s}{D_{ls}} \frac{dp}{dz} dz, \qquad (44)$$

where $dp/dz$ is the redshift probability distribution of source galaxies used for weak lensing measurements. We use the redshift value obtained using EAZY photo-$z$ as described in Section 2.4. In our case, this value comes out to 0.66. This leads to a $\Sigma_{crit}$ value of $4.23 \times 10^{15} M_\odot \text{Mpc}^{-2}$, which is in agreement with the value of $4.55 \times 10^{15} M_\odot \text{Mpc}^{-2}$ obtained by G. Squires et al. (1996). Our value for $\langle D_s/D_{ls} \rangle$ is in agreement with the value of 0.69 obtained by N. Okabe et al. (2010) with images of comparable depth to ours. This is remarkable since both G. Squires et al. (1996) and N. Okabe et al. (2010) were unable to perform photo-$z$ due to lack of color information. A comparison of mass enclosed in the inner part of A2390 with G. Squires et al. (1996) is shown in Figure 13. We find an excellent match up to a radius of 0.7 Mpc from the cluster center. The mismatch after that is likely due to the mass-sheet degeneracy, which states that $\kappa(\theta)$ can be approximately determined to an additive constant. For a detailed discussion on the mass-sheet degeneracy, see M. Bradač et al. (2004).

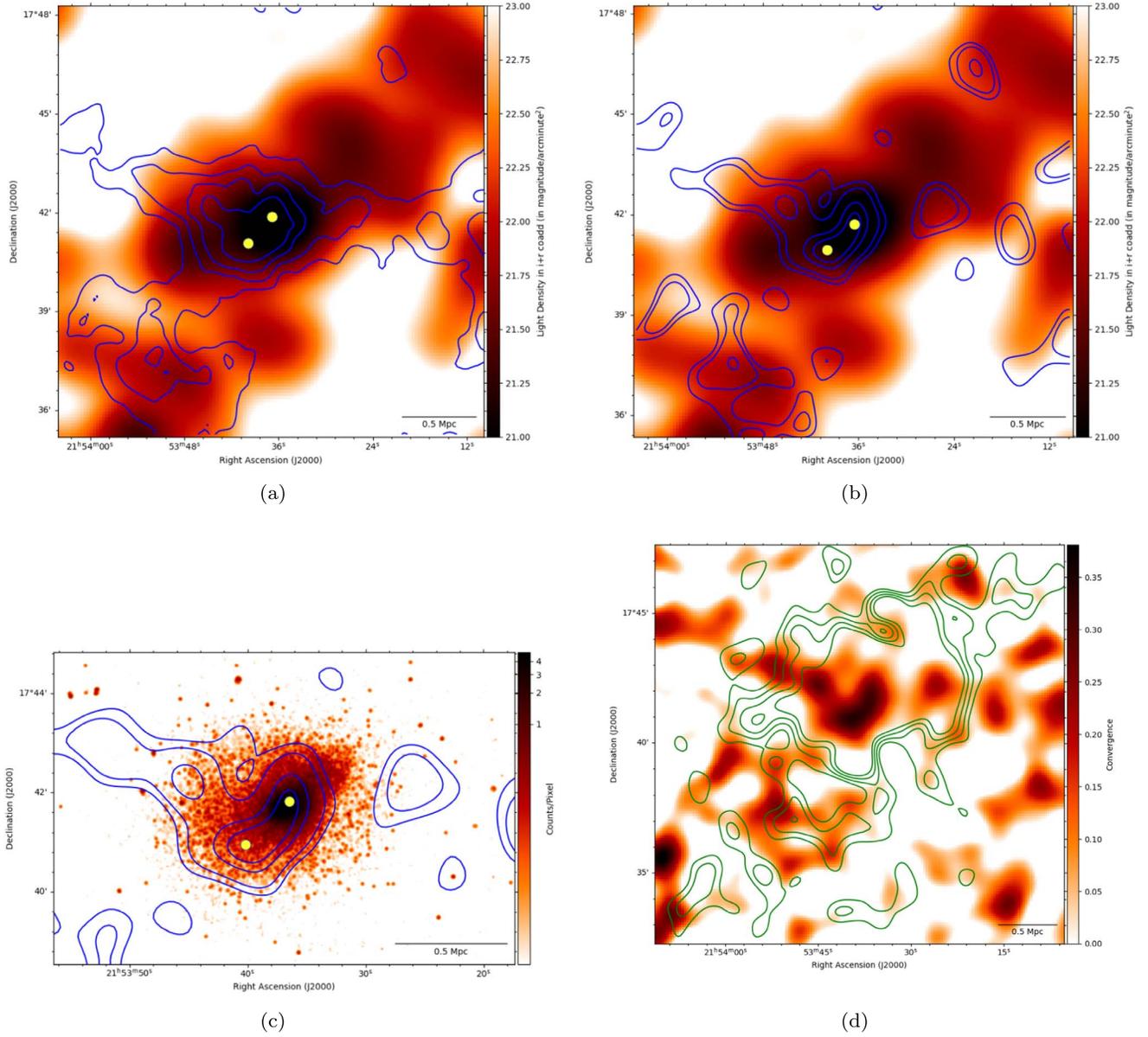

**Figure 10.** (a) The contours of the E-mode (i.e., $M_{ap}$) maps shown in blue are overlaid on the galaxy light density map. Contours are drawn at 0.012, 0.024, 0.036, and 0.048. The light density maps are made by adding light from galaxies in the range $z = 0.15$–$0.55$ weighted by a Gaussian of width $0\farcm73$. The main peak matches up perfectly. The smaller peak on the lower left is distinctly separate from the E-mode peak. The two peaks of the E-mode map are shown as yellow circles. (b) Same as before except the contours are from the convergence map. Contours are drawn at 0.1, 0.15, 0.25, and 0.3. The two peaks of the convergence map are shown as yellow circles. (c) In color shown is the X-ray image from Chandra with the convergence map contours overlaid. A slight swirl pattern in the X-ray can be seen matching the swirl pattern of the convergence contours. (d) The convergence map is shown as background. Contours show the compact-component-subtracted radio images from LOFAR obtained after careful reanalysis of DR2 data by M. Cianfaglione et al. (2024, in preparation). Contours are drawn at 0.0014, 0.0021, 0.0028, and 0.0035 Jy beam$^{-1}$, which corresponds to 2, 3, 4, and $5\sigma$ significance, respectively.

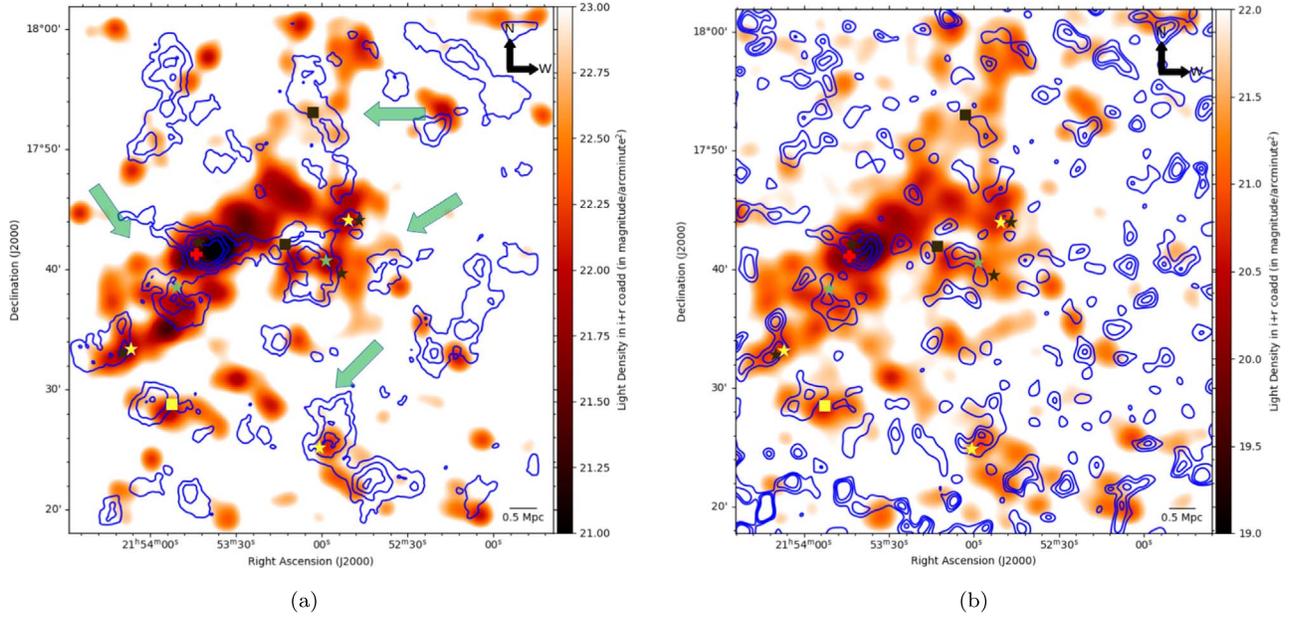

**Figure 11.** (a) The contours show the E-mode (i.e., $M_{ap}$) maps constructed from the galaxy sample in the range $z = 0.4$–$2.0$ overlaid on the light density map of galaxies in the range $z = 0.15$–$0.55$. We fixed $\sigma_{M_{ap}} = 3.5 \times 10^{-4} \pm 1.75 \times 10^{-5}$. Contours are drawn at levels of 0.012, 0.024, 0.036, and 0.048. The background is a light density map that was smoothed with a Gaussian width of $0.'73$. Four major filamentary structures connecting the galaxy cluster are visible and are shown in green arrows: one to the north, one to the west (right), and one toward the southwest (down right); another structure to the southeast is also visible. The filaments seen from the contours broadly agrees with the filaments seen in the light density maps. The location clusters obtained from NED are shown as symbols. The squares are obtained from X-ray studies while the stars are obtained from optical data. Cross symbols show cluster locations obtained using the SZ effect on Planck data. Each symbol and the corresponding literature is discussed in detail in Section 6. (b) Contours of $\kappa(\theta)$ overlaid on the light density maps smoothed with a Gaussian of $0.'73$ of galaxies in the range $z = 0.15$–$0.55$. Contours have been drawn at levels of 0.1, 0.2, and 0.3. The $\kappa(\theta)$ map has been smoothed with a Gaussian of width $16.''5$. The bimodal mass distribution of the central region of A2390 can be clearly seen at R.A. $= 21^h:53^m:36.^s8$ and decl. $= 17°: 41': 43''$. The other peaks are consistent with the mass concentrations shown by E-mode maps. We find also increased noise at the edges of the $\kappa(\theta)$ map.

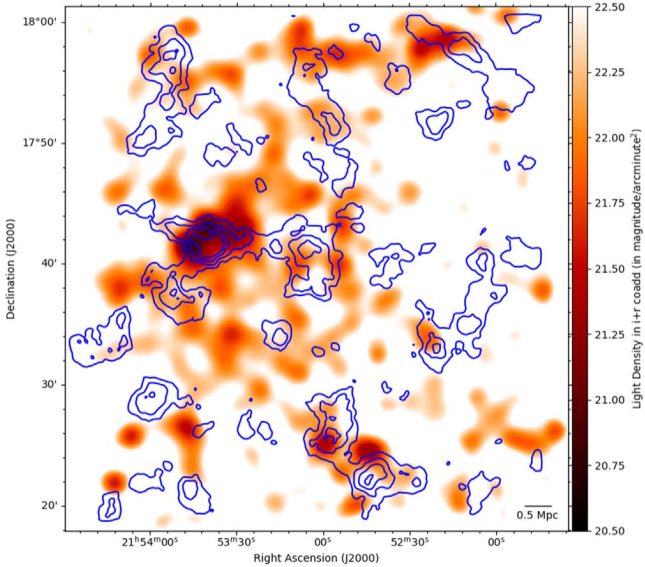

**Figure 12.** The contours show the E-mode (i.e., $M_{ap}$) maps constructed from the galaxy sample in the range $z = 0.4$–$2.0$ overlaid on the light density map from galaxies in the redshift range 0.35–0.75. Contours are drawn at levels of 0.012, 0.024, 0.036, and 0.048. We fixed $\sigma_{M_{ap}} = 3.5 \times 10^{-4} \pm 1.75 \times 10^{-5}$. The background is a light density map that was smoothed with a Gaussian width of $0.'73$.

### 4.3. 3D Slices

The ability to make slices of mass distribution at various redshifts would be the ultimate test of any deep weak lensing analysis. One of the first maps of 3D weak lensing reconstruction was done by R. Massey et al. (2007). It used high-quality images obtained from the Hubble Space Telescope to measure shear in each redshift bin. However, such reconstructions are extremely rare in the literature, especially with ground-based data. One of the challenges in making such a 3D map is galaxy density. In weak lensing one typically requires a few tens of galaxies per square arcminute to measure shear. However, when these are divided into redshift bins the galaxy density is drastically decreased, which introduces more noise. Below, we describe our approach.

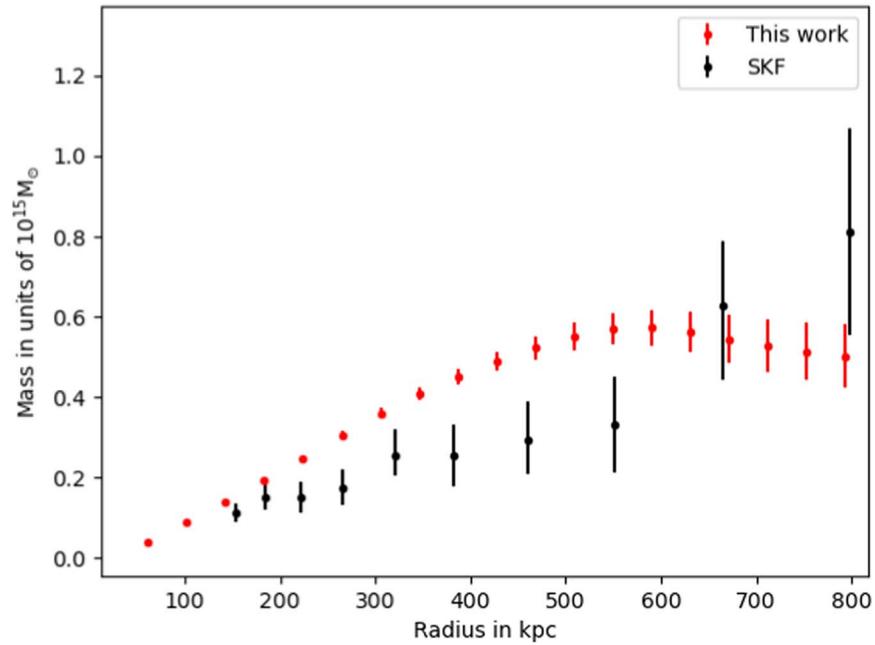

**Figure 13.** Enclosed mass as a function of radius from the center of the galaxy cluster. The center is located on the BCG at R.A. = 21$^h$:53$^m$:36$^s$.8 and decl. = 17°:41′:43″. We used Poisson error bars and multiplied them with a factor of 10, as found in PhoSim simulations, to take into account systematic/non-Poisson errors. We find our data match extremely well with G. Squires et al. (1996) up to a radius of 700 kpc. The mismatch beyond that is likely due to the mass-sheet degeneracy and can be remedied using a small additive constant.

We construct an E-mode map, $E_1$, with all galaxies in the redshift range $z = 0.01$–2.0. We only consider the sources that satisfy the conditions mentioned in Section 4.1. This map, in an ideal case, traces most cosmic structures from $z = 0$ to 2.0. We construct a second E-mode map, $E_2$, with all galaxies in the range $z = 0.5$–2.0. This E-mode map boosts the structures in the range $z = 0.01$–0.5 since it only contains galaxies background to $z = 0.5$. This can be written down as

$$E_1 = S_1\left(\frac{N_2}{N_1 + N_2}\right) + S_2, \quad (45)$$

$$E_2 = S_1 + S_2, \quad (46)$$

where $S_1$ is the cosmic structure is the foreground slice and $S_2$ is the structure in the background slice. $N_1$ is the number of sources in the redshift range $z = 0.01$–0.5, i.e., the foreground slice, and $N_2$ is the number of sources in the redshift range $z = 0.5$–2.0, i.e., the background slice. The factor $\left(\frac{N_2}{N_1 + N_2}\right)$ signifies that only the population $N_2$ traces structure $S_1$.

It was found significant noise arises from the fact that the resolution of E-mode maps is different, with $E_1$ having significantly better resolution than $E_2$. This is because the number of galaxies used to construct $E_2$ is approximately half of the number of galaxies used to construct $E_1$. To remedy this we bring all E-mode maps to the same resolution. We smooth $E_1$ by a Gaussian of width 4 pixels, i.e., 22″, and $E_2$ by a Gaussian of width 2 pixels, i.e., 11″. This was visually determined after careful examination of the two E-mode images, i.e., $E_1$ and $E_2$.

While in theory all pixels of the E-mode maps matter, we note that in reality the shear signals are extremely small. Hence, only the brightest few pixels in the E-mode image contain information. It was found that capping the minimum value of E-mode maps at one standard deviation from the media of all pixels in the image produces better results. This value in our case comes out to approximately 0.014 for $E_1$ and 0.018 for $E_2$. The E-mode maps were produced in an adaptive manner constraining $\sigma_{M_{ap}} = 4.5 \times 10^{-4} \pm 2.25 \times 10^{-5}$. Next, using inversion of Equations (45) and (46) mentioned above, we recovered the two slices. The E-mode contours of both the foreground and background slice are shown in Figure 14. We find the foreground slice contains signal from A2390 while the background slice is completely devoid of signal from A2390. A few of the more prominent substructures can be seen in the foreground slice. The background slice is very noisy, and the contours do not seem well correlated to light density maps. This is not surprising since after all the cuts mentioned above $E_2$ had an extremely low source density of 8 galaxies arcmin$^{-2}$.

## 5. Discussion

A2390 has been extensively studied in X-ray (M. Pierre et al. 1996; S. Allen et al. 2001; R. Martino et al. 2014; S. S. Sonkamble et al. 2015), optical (R. G. Abraham et al. 1996; J. B. Hutchings et al. 2002), radio (M. Bacchi et al. 2003; P. Augusto et al. 2006; M. W. Sommer et al. 2016; F. Savini et al. 2019), weak lensing (G. Squires et al. 1996; K. Umetsu et al. 2009; N. Okabe et al. 2010; A. von der Linden et al. 2014), and strong lensing studies (D. Narasimha & S. M. Chitre 1993). The central continuum source of the brightest cluster galaxy (BCG) shows young, compact, and self-absorbed jets (A. C. Edge et al. 1999; P. Augusto et al.

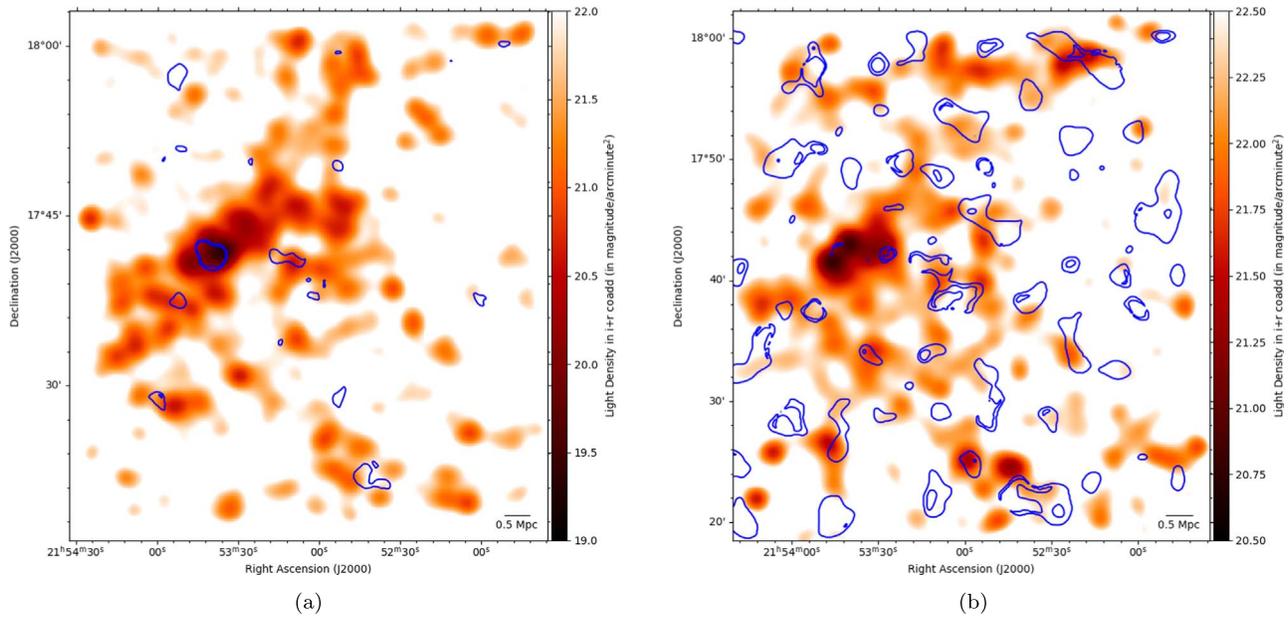

**Figure 14.** (a) The adaptive E-mode contours from the foreground slice ($z = 0.01–0.6$) overlaid on an image showing the light density of galaxies in the redshift range $z = 0.15–0.55$ smoothed with a Gaussian of width $0.'73$. (b) The contours depicted are from the adaptive E-mode maps of the background slice ($z = 0.5–2$) overlaid on an image showing the light density of galaxies in the redshift range $z = 0.35–0.75$ similarly smoothed. In both cases contours are placed at levels of 0.02, 0.07, and 0.12. There is significant noise in our maps. The signal from A2390 appears only in the foreground slice as expected.

2006). Accretion toward the central supermassive black hole has also been inferred from CO, CN, SiO, HCN, and HCO+ absorption lines against the radio continuum source, which show molecular gas clouds moving in toward the galaxy center at roughly 100–300 km s$^{-1}$ (T. Rose et al. 2019, 2024b). In Figure 15, we show a composite image of X-ray (in red) and the convergence map (in blue) overlaid on the optical image of A2390. In the following sections, we briefly discuss these previous findings and compare them to our results.

### 5.1. X-Ray

High-resolution X-ray images of A2390 were obtained and analyzed by S. Allen et al. (2001). ROSAT data for this cluster have been analyzed by M. Pierre et al. (1996). However, the Chandra data are much more detailed, and hence we focus on these. S. Allen et al. (2001) found that the X-ray profile can be fit well to a NFW profile. A variety of other profiles, such as a softened isothermal sphere and a full isothermal sphere, were also found to provide reasonably good fits. They also note that this cluster does not seem to be completely relaxed. Both S. Allen et al. (2001) and T. Rose et al. (2024a) find evidence of excess X-ray emission approximately 5″ southeast of the central peak, which coincides with the BCG. We note here that this excess was found using two independent methods. S. Allen et al. 2001 used adaptive smoothing to find the excess, while T. Rose et al. (2024a) subtracted a double beta model. Both methods show the excess emission is approximately 20 kpc (5″) from the central peak. It is hypothesized by S. Allen et al. (2001) that the cluster has not fully relaxed from the last merger. The model-subtracted images of T. Rose et al. (2024a) are interesting because they seem to show two cores orbiting one another (their Figure 6). However, there is a chance the excesses and depressions seen in the model-subtracted images are a result of AGN activity. Indeed, this has been suggested by S. S. Sonkamble et al. (2015). Evidence of current AGN activity at kiloparsec scales was found by P. Augusto et al.

(2006), with moderate high radio-frequency variability of the radio continuum since 2015 also identified by T. Rose et al. (2022).

### 5.2. Radio Observations

It has been known for decades that massive merging clusters are likely to host radio emission in the form of radio halos and radio relics (R. J. van Weeren et al. 2019). The synchrotron radio emission would be caused by the reacceleration of cosmic-ray electrons by turbulent motions that develop in the ICM during cluster mergers.

In the past years, radio observations have started to reveal radio halos in clusters that are not undergoing major mergers, and that—in some cases—host a cool core (e.g., A. Bonafede et al. 2014; M. W. Sommer et al. 2016; T. Venturi et al. 2017; F. Savini et al. 2019; N. Biava et al. 2024). These results indicate that these sources might be connected to the occurrence of minor/off-axis mergers, though it remains unclear how minor mergers could initiate continuum emission on megaparsec scales.

A2390 is one of these clusters, as it hosts both a cool core and signs of minor dynamical disturbances from the X-ray morphological parameters. Diffuse radio emission in A2390 was first discovered and classified as a radio mini-halo by M. Bacchi et al. (2003) and then as a radio halo by M. W. Sommer et al. (2016). LOFAR observations revealed the presence of a double radio galaxy with the lobes extending in the east–west axis for ~600 kpc, and F. Savini et al. (2019) could not distinguish the radio emission from the radio galaxy from a possible contribution from the radio halo, leaving the possibility of a radio halo open. F. Savini et al. (2019) also noted that radio galaxies of such a size are uncommon at the center of galaxy clusters, as the ICM prevents the expansion of the lobes to such large scales. LOFAR observations of A2390 have also been published in the LOFAR Data Release 2

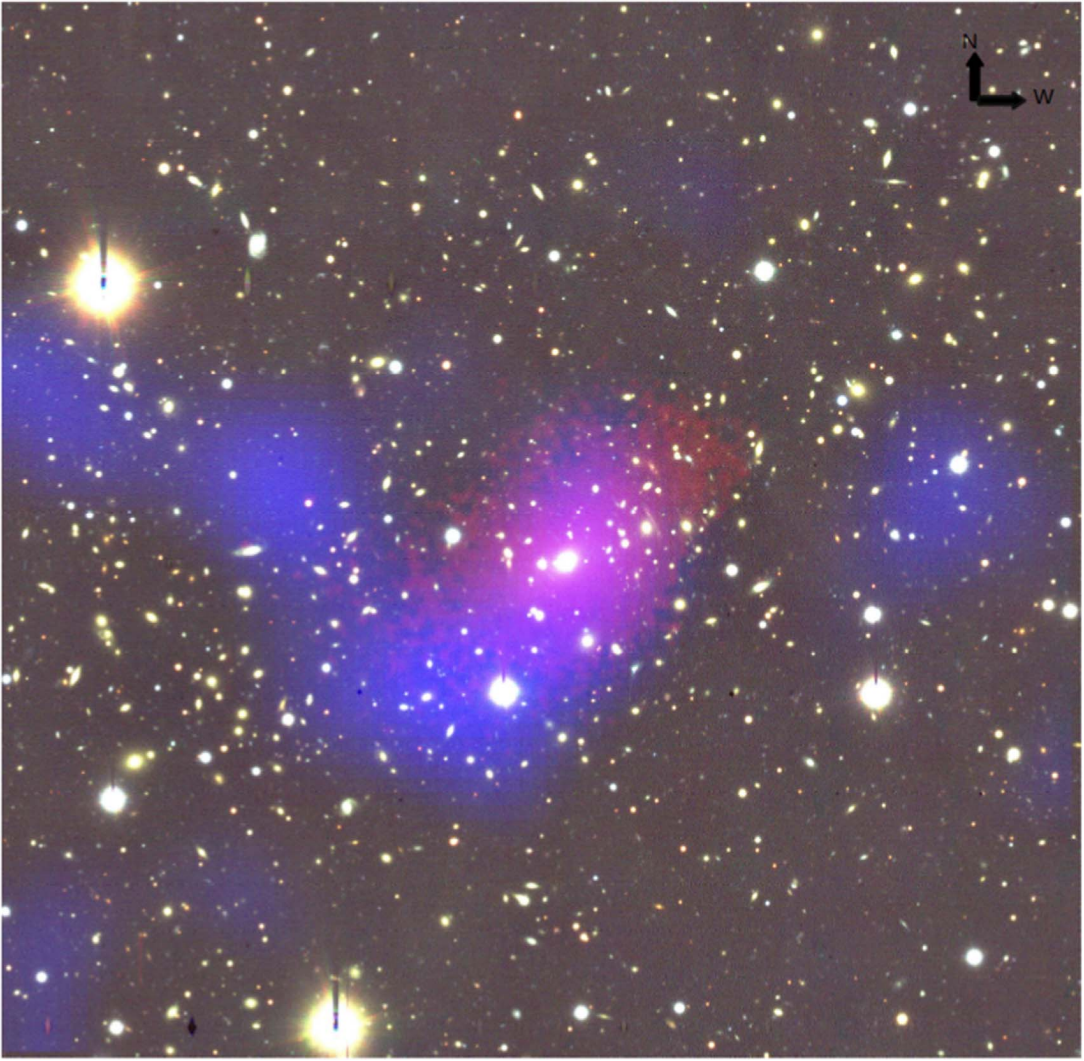

**Figure 15.** A composite image of A2390. The X-ray obtained using Chandra (S. Allen et al. 2001) is shown in red and our mass map (convergence) in blue. It is overlaid on the RGB color image of A2390 created using a *g*-band coadd for blue, *r*-band coadd for green, and *i*-band coadd for red. Optical images have been obtained using WIYN-ODI. At the center of the image, the bright elliptical galaxy is the BCG. The X-ray shows a swirl-like pattern in the same direction as the DM.

(A. Botteon et al. 2022), and the authors concluded that most of the emission came from the radio galaxy.

M. Cianfaglione et al. (2024, in preparation) have reanalyzed the data from the LOFAR Data Release 2 (DR2) and subtracted the central AGN using an approach tuned "ad hoc" for this cluster. Specifically, they subtracted all the emission on scales smaller than $375''$, corresponding to 1390 kpc, and reimaged the data at low resolution (that is, $1'$) to gain sensitivity toward the extended emission (Figure 10(d)). They found residual radio emission with a flux density $S(144\,\mathrm{MHz}) = 0.16 \pm 0.03$ Jy, corresponding to a monochromatic power $P(144\,\mathrm{MHz}) = 2.7 \pm 0.5 \times 10^{25}$ W Hz$^{-1}$. In addition, the emission radial profile follows the exponential profile typically found in radio halos. The radio halo emission extends for 1450 kpc. As the size, power, and radio profile are all in line with those of radio halos in clusters of similar mass (V. Cuciti et al. 2023), they concluded that the residual emission is actually a radio halo. The presence of a radio halo reinforces the results obtained in this work, as they are in line with the presence of dynamical activity. If the merger is indeed in a late stage, as derived in this work, we expect the radio halo emission to be steep (G. Brunetti & T. W. Jones 2014; F. Savini et al. 2019; N. Biava et al. 2024). M. W. Sommer et al. (2016) using Very Large Array (VLA) data estimates the overall spectral index to be 1.6, which is fairly high.

High-resolution radio images of the BCG using the Atacama Large Millimeter/submillimeter Array (ALMA) to detect CO reveal an extended tail (T. Rose et al. 2024a). They considered various scenarios, including both outflows and inflows, and came to the conclusion that a recent gravitational disturbance of the central BCG is the most likely explanation of the extended tail. L. Y. Alcorn et al. (2023) using Canada–France–Hawaii Telescope (CFHT)/SITELLE finds the BCG has a tail, matching the findings of T. Rose et al. (2024a). All of this points to a fairly disturbed structure in the central region.

### 5.3. Lensing and Spectroscopy

R. G. Abraham et al. (1996) made spectroscopic measurements of this galaxy cluster using CFHT. They found spectroscopic evidence of an infalling galaxy stream from the northwest corner approximately $10'$ from the center of A2390. This can also be seen from the light density maps shown in Figure 11(a). This group of galaxies has also been identified using optical cluster-finding algorithms, shown as yellow and black stars in Figure 11(a).

A strong lensing study of this cluster was performed by D. Narasimha & S. M. Chitre (1993). Weak lensing studies have also been conducted using data obtained from CFHT (G. Squires et al. 1996) and the Subaru Telescope (K. Umetsu et al. 2009; N. Okabe et al. 2010). It was also studied by A. von der Linden et al. (2014) by using data from both telescopes. These studies, with the exception of A. von der Linden et al. (2014), do not find evidence of a merger in the galaxy cluster using weak lensing maps. However, we note here that the only two studies with depth comparable to ours are N. Okabe et al. (2010) and A. von der Linden et al. (2014). We also note A. von der Linden et al. (2014) come to the conclusion that evidence of merger activity from the northwest, presumably the same galaxy group discovered by R. G. Abraham et al. (1996), is present in this cluster. While our light maps in Figure 11(a) clearly show this, we are not able to recover this with very high confidence. We detect the peak of this smaller group with somewhat lower confidence. We believe this is a minor and early-stage merger event. In addition to this, we also find evidence of gravitational disturbance in the central region of A2390. Our convergence maps show the DM cores are in the process of merging. The aperture mass maps of A. von der Linden et al. (2014) look similar to ours, in the sense that they find a bimodal mass distribution at the core. Due to the significantly worse resolution of their map, unfortunately any further comparison is difficult.

### 5.4. Late-merger Hypothesis

We believe A2390 is a case of an extreme late-stage merger as suggested by S. Allen et al. (2001). This seems to be the most likely explanation of all observed data. During this merger, the hot gas experienced friction when orbiting the cluster, while the DM component experienced very little of this friction, if any. This caused the hot gas to lose angular momentum and fall into the central core faster than the DM. Currently, the hot gas is almost merged with the central core since it can only be detected using adaptive smoothing (S. Allen et al. 2001) or model subtraction (T. Rose et al. 2024a). The DM core still shows signs of an ongoing merger. In our mass maps, the mass excess is found southeast of the main peak at a separation of approximately $1'$ or 220 kpc. This separation, while in the exact same direction as the X-ray excess, is about 10 times more when compared to the X-ray separation of $5''$, i.e., 20 kpc. A slight swirl pattern in the lower left of the X-ray image in Figure 10(c) is visible and matches up with the swirl seen on the overlaid convergence contours, lending support to this hypothesis.

The late-stage merger would cause gravitational disturbance primarily in the central regions of the cluster. This is supported by the extended tail of the BCG found by T. Rose et al. (2024a). The direction of the tail also matches with the expected direction from a late-stage merger. The process of an ongoing merger from the southeast would cause the plume in the BCG to be in the northwest direction.

The extended radio emission found by M. W. Sommer et al. (2016) and M. Cianfaglione et al. (2024, in preparation) also supports our hypothesis. The late-stage orbital motion of two heavy DM cores will introduce significant turbulence and ripples in the ICM, which leads to the re-energizing of leptons, which in turn leads to a radio halo with a steep spectral index. M. W. Sommer et al. (2016) estimated the spectral index to 1.6, which is very steep and consistent with our hypothesis of a late-stage merger.

### 6. Smaller Groups and Large-scale Structure

Weak lensing provides one of the most powerful ways to map large-scale structures of DM through the Universe (A. Refregier 2003). This is especially useful in probing regions of lower density such as filamentary structures and smaller galaxy groups since shear depends linearly on mass enclosed. Other methods such as X-rays are ineffective in studying these regions since X-ray brightness depends on density squared (S. Ettori 2000). Hence, X-ray signals from low-density regions cannot be detected. These regions are difficult to detect in radio wavelengths as well due to lower concentrations of ICM and ultrarelativistic particles in the ICM. Reliable detection of these lower-mass structures can be used in conjunction with the mass, redshift, and distribution of larger structures, such as galaxy clusters, to better constrain cosmological models (F. Bernardeau et al. 1997; B. Jain & U. Seljak 1997).

In Figure 11(a), we show the light density of the $i+r$ coadd as the background color. We only consider galaxies in the redshift range $z = 0.15$–$0.55$ and the galaxies that pass all the size, flag, and flux cuts mentioned in Section 4.1. This is the baryonic mass distribution and is roughly expected to follow the large-scale structure. The contours from the aperture mass map are overlaid on this. The symbols show the reported galaxy groups and clusters on NED. The star symbols were obtained using cluster-finding methods that rely on optical data. The yellow star corresponds to the list published by Z. L. Wen et al. (2012) and later updated by Z. L. Wen & J. L. Han (2015). The black star corresponds to the list published by E. Rozo et al. (2015). Both of these methods primarily use SDSS data. The green star corresponds to the list published by R. R. Gal et al. (2009) using data from the Digitized Second Palomar Observatory Sky Survey. X-ray analysis of this region of the sky is primarily based on XMM-Newton data. The locations of smaller groups and clusters based on available X-ray data are shown as squares. The black square shows the location of groups mentioned in C. P. Haines et al. (2018). The yellow square corresponds to locations mentioned in P. A. Giles et al. (2022). The red cross is the cluster location derived from the SZ effect in Planck data (P. A. R. Ade et al. 2016; R. Khatri 2016). It is clear in Figure 11(a) that most of the smaller galaxy groups are successfully detected by our aperture mass map. There are some regions in our map that have high light density but are neither detected by weak lensing nor the different galaxy group detection methods mentioned above. This could arise from a variety of factors. The light maps could be contaminated with light from sources that are not within the desired redshift range. It could also be the case that our images are significantly deeper than any existing data to which the cluster-finding algorithms have been applied. Thus, these structures are visible in our light map only.

It has been shown that aperture mass statistics or significance maps are able to trace the cosmic filamentary structures (M. Jauzac et al. 2012; K. HyeongHan et al. 2024). We are able to recover a few broad filamentary structures as shown in Figure 11(a). We notice four major structures in the light density maps: one toward the north, one to the west, one to the southwest, and one to the southeast. All of these are broadly traced by the contours. The contours also trace a structure to the northeast where there seems to be a small galaxy group. We note that neither the aperture mass map nor the convergence map is able to detect the filament northwest of the central region. In Figure 12, we overlay the E-mode contours on light density maps of galaxies in the redshift range $z = 0.35$–$0.75$. The light density maps are made using the same method as before.

## 7. Conclusion

In this paper, we present a weak lensing analysis of the galaxy cluster A2390 using extremely deep images obtained from WIYN-ODI. We introduce a novel method that allows us to obtain shear information from galaxies which are measured to be smaller than the PSF. This allows us to create mass maps with higher source density than previously possible with images of similar depth. We measured shapes in individual exposures using a moment-matching algorithm. The forced measurement method was used when the SNR was too low for convergence. The aperture mass maps obtained show that we are able to recover most of the smaller galaxy groups, identified using optical and X-ray cluster-finding algorithms. A group of galaxies approximately $10'$ northwest of the cluster was also found and appears to be in the process of infalling into the cluster from spectroscopic data. In addition to this, most of the filamentary structures around A2390 were also recovered in the aperture mass maps. Within a radius of $\sim 220$ kpc from the BCG, A2390 appears to have a bimodal mass distribution, with the smaller peak being southeast of the main peak, which is consistent with the X-ray excess found in Chandra data. The separation between the peaks is 220 kpc, whereas the separation in X-ray is 20 kpc. This suggests that A2390 is a case of an extreme late-stage merger, with the hot gas close to the center now being completely relaxed following the most recent merger. However, due to the lack of friction during the infall period, the DM is still actively merging with the main DM core. The merger hypothesis is supported by CO radio observations using ALMA, which find a tail in the BCG that can be explained by a recent gravitational disturbance. The high spectral index of 1.6 found using VLA data and the radio halo found in both VLA and careful re-examination of LOFAR DR2 also support this hypothesis. If true, more such findings in other galaxy clusters along with simulations might help us put upper limits on the cross section of DM. We find our mass estimate for this galaxy cluster is consistent with previously published results.

## Acknowledgments

The authors thank the anonymous referee for the useful comments and suggestions. The authors would like to thank Purdue University for its continued support. The authors would also like to thank the AAS Publication Support Fund for financial support of this article. We are also very grateful to the WIYN-ODI PPA team, especially Wilson Liu, Nick Smith, Arvind Gopu, and all the telescope operators for their help in obtaining excellent-quality data. We also thank the Purdue Rosen Center for Advanced Computing (RCAC) for access to computing facilities that have been extensively used in this paper.

The data analysis was done with python, and the authors acknowledge the use of astropy (Astropy Collaboration et al. 2013, 2018, 2022), numpy (C. R. Harris et al. 2020), scipy (P. Virtanen et al. 2020), matplotlib (J. D. Hunter 2007), and aplpy (T. Robitaille & E. Bressert 2012; T. Robitaille 2019). This research has made use of the NASA/IPAC Extragalactic Database (NED), which is funded by the National Aeronautics and Space Administration and operated by the California Institute of Technology.

*Facility:* WIYN.

## ORCID iDs

A. Dutta https://orcid.org/0009-0000-1088-4653
J. R. Peterson https://orcid.org/0000-0001-5471-9609
A. Bonafede https://orcid.org/0000-0002-5068-4581